\documentclass[reprint,amsmath,amssymb,aps]{revtex4-2}
\usepackage[colorlinks,citecolor=blue,linkcolor=blue,anchorcolor=blue,filecolor=blue,
urlcolor=blue]{hyperref}
\usepackage{graphicx}
\usepackage{ulem}
\usepackage{dcolumn}
\usepackage{bm}
\usepackage{todonotes}

\begin{document}

\title{Pseudoscalar mesons from a PNJL model at zero temperature}

\author{R. M. Aguirre$^1$ and O. Louren\c{c}o$^{2,3}$}
\affiliation{ \mbox{$^1$Departamento de Fisica, Facultad de
Ciencias Exactas, Universidad Nacional de La Plata,} \mbox{and
IFLP, UNLP-CONICET, C.C. 67 (1900) La Plata, Argentina}
\\
\mbox{$^2$Departamento de F\'isica, Instituto Tecnol\'ogico de
Aeron\'autica, DCTA, 12228-900, S\~ao Jos\'e dos Campos, SP,
Brazil}
\\
\mbox{$^3$Universit\'e de Lyon, Universit\'e Claude Bernard Lyon
1, CNRS/IN2P3, IP2I Lyon, UMR 5822, F-69622, Villeurbanne, France}
}

\date{\today}

\begin{abstract}
We study pseudoscalar $\pi$, $K$ and $\eta$ meson properties, such
as masses and couplings, in dense matter at zero temperature. We
use a recently proposed phenomenological quark model, known as the
PNJL0, which takes into account the confinement/deconfinement
phase transition by means of the traced Polyakov loop ($\Phi$)
which serves as an order parameter at zero temperature. We
consider two different scenarios, namely, symmetric quark matter
with equal chemical potentials for all the flavors, and the beta
equilibrated matter. In the latter case the hadron-quark phase
transition is implemented by a two model approach. For the hadron
side we use a relativistic mean-field model with density dependent
couplings.  We show that $\Phi$ induces abrupt changes in the
mesons properties with gap sizes regulated by the phenomenological
gluonic sector of the model.
\end{abstract}

\maketitle

\section{Introduction}

The study of meson properties under extreme conditions is expected
to reveal significant features of the Quantum Chromodynamics (QCD)
phase diagram~\cite{hayano}. For instance, the restoration of
chiral symmetry at high temperatures would manifest by the
degeneracy of the masses of the chiral partners $\sigma-\pi$ and a
drastic drop of the effective coupling between pion and
quarks~\cite{HANSEN,FU_LIU}. An interesting outcome is the
prediction that mesonic excitations exist beyond the critical
temperature for deconfinement and even for chiral symmetry
restoration~\cite{HANSEN,FU_LIU,ABUKI,YAMAZAKI,COSTA2,GUPTA,RATTI}.
The enhanced fraction of bound states of strange quarks observed
at high energies in heavy collisions at the Large Hadron Collider
would give the experimental evidence.  The situation may be
consistently interpreted as a flavor staggered
deconfinement~\cite{BELLWIED}. This issue has been intensively
studied for non-zero temperatures, leading to the conclusion that
a coexistence of confined and deconfined
phases~\cite{BUGAEV,DITORO} is feasible. The simulations performed
in~\cite{DITORO} have shown that the collision of stable nuclei
could produce a mixed phase at densities 3-4 times the normal
nuclear density and temperatures around 50 MeV.

There are physical systems, as in the core of massive neutron
stars, for which the coexistence of phases at nearly zero
temperature is a plausible scenario~\cite{GLENDENNING}. The pure
quark matter phase is also another possibility in this particular
system~\cite{nat1,nat2}. The relevance of the $T=0$ and high
density regime has been remarked by the finding of new phases in
the QCD diagram, as for instance the kaon condensate for different
flavor compositions~\cite{BARDUCCI}, or in a superconducting quark
matter state~\cite{SCHAFER,BASLER}.

The usual way to deal with a mixed phase is to use a dual scheme,
taking hadrons as effective degrees of freedom below the threshold
transition and a model of deconfined quarks above it. These
different descriptions are combined in a continuous manner or not,
depending on the nature of the transition. Furthermore, some
hybrid models has been proposed that combine in the same framework
bound states and deconfined quarks, as for
instance~\cite{STEINHEIMER,BENIC}. In~\cite{STEINHEIMER} the
Polyakov loop is  explicitly used, whereas~\cite{BENIC} modifies
the Fermi statistical distribution functions by restricting the
momentum regime accessible for quarks or hadrons. By its very
construction, the Polyakov loop has been used only for finite
temperature situations. It has been recently proposed in
Refs.~\cite{MATTOS,Mattos_2019,Mattos_2021} an extension of the
Nambu-Jona Lasinio model (NJL) at zero temperature, with
interacting quarks coupled to a scalar background field $\Phi$
that effectively mimics the quark-gluon dynamics, namely, the
transition from strong to weak regimes of the quark-quark
interaction. In such a model, called as PNJL0 model, it is
possible to clearly determine different thermodynamical phases in
which quarks are treated as confined or deconfined particles.
Furthermore, the particular phase in which the system presents
restored chiral symmetry, but with confined quarks, can also be
identified. The back reaction between quark and gluon sectors is
also a feature of the PNJL0 model. In Ref.~\cite{Mattos_2021} the
thermodynamics of the 2-flavor version of the model was studied,
and in Ref.~\cite{MATTOS} the extension to the SU(3) system was
implemented with some applications to the stellar matter. More
specifically, the construction of hybrid stars was analyzed in
which the PNJL0 model was coupled to a hadronic relativistic
mean-field~(RMF) model.

Different studies have shown that the modification of the meson
properties in a dense medium has significant consequences for a
wide range of phenomena. In the astronomical scale, the
condensation of pions and kaons in  stars~\cite{PONS,VIJAYAN} is a
longstanding issue due to its influence on the cooling
process~\cite{YAKOVLEV} or the pulsar glitches. While in the
microscopic scale, the phenomenology of mesic
nuclei~\cite{FRIEDMAN,MUTO,CIEPLY,SONG,NAGAHIRO,NAGAHIRO2,JIDO}
has revealed some supporting evidence, as for instance the
systematics of pionic nuclei which suggests a decrease of the pion
decay constant~\cite{FRIEDMAN}. Furthermore, calculations of a
density dependent polarization insertion within a chiral inspired
parametrization of hadronic masses or vertices predict stable
configurations of nuclear bound states of
$\eta$~\cite{CIEPLY,NAGAHIRO} as well as of $\eta'$
mesons~\cite{NAGAHIRO2}.

The combination of first principles and such phenomenological
issues has given different theoretical pictures of the interaction
between nucleons and the pseudoscalar mesons. Besides the well
constrained nucleon-pion interaction, there is a variety of kaon
models~\cite{TOLOS}, as for instance effective
kaon-meson~\cite{GLENDENNING2} and chiral symmetry inspired
kaon-nucleon~\cite{MISHRA_SANYAL,MARUYAMA} vertices. The dynamics
of the $\eta-\eta'$ system in a dense environment is less known,
and its theoretical description has been mainly based on chiral
constructions~\cite{ZHONG,KUMAR}, or relies on covariant hadronic
field models~\cite{SONG,JIDO}. In this paper, we calculate the
meson properties by using the PNJL0 model and a particular RMF
hadronic one, both described in Sec.~\ref{form}. In
Sec.~\ref{mesonprop} we present the theoretical framework for the
evaluation of the meson polarization for both quark and hadronic
models. The results for the masses and effective couplings of
$\pi$, $K$, $\eta$ mesons are shown and discussed in
Sec.~\ref{RESULTS}  with special attention to the
confinement/deconfinement phase transition. Finally, the summary
and concluding remarks are presented in Sec.~\ref{summary}.

\section{PNJL0 model: symmetric and stellar matter cases}
\label{form}

The Lagrangian density that describes the three-flavor version of
the PNJL model with scalar, vector, and 't~Hoof channels is given
by~\cite{MATTOS}
\begin{align}
\mathcal{L}_{\mbox{\tiny PNJL}} &= \bar{q}(i\gamma_\mu D^\mu -
\hat{m})q  - \mathcal{U}(\Phi,\Phi ^*,T)
\nonumber\\
&+\frac{G_s}{2}\sum_{a=0}^{8}\left[(\bar{q}\lambda_aq)^2-(\bar{q}\gamma_5
\lambda_aq)^2 \right]
\nonumber\\
&-\frac{G_V}{2}\sum_{a=0}^{8}\left[(\bar{q}\gamma_\mu\lambda_aq)^2+(\bar{q}\gamma_\mu\gamma_5
\lambda_aq)^2 \right]
\nonumber\\
&+ K[\mbox{det}_f(\bar{q}(1-\gamma_5)q) +
\mbox{det}_f(\bar{q}(1+\gamma_5)q)], \label{dlpnjl}
\end{align}
where $q$ is a vector of three spinors $q_f$ for $f=u,d,s$,
$D^\mu\equiv\partial^\mu+iA^\mu $ with $A^\mu=\delta^\mu_0A_0$ and
$A_0=gA_a^0\lambda_a / 2$ ($g$ is the gauge coupling),
$\hat{m}=\mbox{diag}(m_u,m_d,m_s)$ is a matrix of current quark
masses in flavor space, and $\lambda_a$ are the SU(3) Gell-Mann
matrices. The strength of scalar, vector, and 't~Hoof interactions
is determined by the constants $G_s$, $G_V$, and $K$,
respectively. The use of the mean-field approximation leads this
quantity to
\begin{align}
\mathcal{L}_{\mbox{\tiny PNJL}} &= \sum_f\bar{q}_f(i\gamma_\mu
D^\mu - M_f)q_f -G_s\sum_f{\rho_{sf}^2}
\nonumber\\
&+ G_V\sum_f\rho_f^2 - 4K\prod_f\rho_{sf} -
\mathcal{U}(\Phi,\Phi^*,T). \label{dlpnjlmfa}
\end{align}
Notice the inclusion of the Polyakov potential
$\mathcal{U}(\Phi,\Phi ^*,T)$ that depends on the traced Polyakov
loop, defined as
\begin{align}
\Phi&=\frac{1}{3}\rm{Tr}\left[\,\,\rm{exp}\left(i\int_0^{1/T}d\tau\,A_4\right)\right]
\nonumber \\
&=
\frac{1}{3}\left[\rm{e}^{i(\phi_3+\phi_8/\sqrt{3})}+\rm{e}^{i(-\phi_3+\phi_8/\sqrt{3})}
+\rm{e}^{-2i\phi_8/\sqrt{3}}\right], \label{traced}
\end{align}
written in the Polyakov gauge where
$\phi=\phi_3\lambda_3+\phi_8\lambda_8$, and $A_4=iA_0\equiv
T\phi$. $\Phi^*$ is the conjugate of $\Phi$.

In general, the Polyakov potential is zero at $T=0$, as one can
see for instance in
Refs.~\cite{ratti_weise,ratti_weise2,fukushima,rossner}.
Therefore,  all thermodynamical quantities related to the PNJL
model reduces to the NJL model ones at this regime, and it is not
possible to investigate deconfinement effects at this level. In
order to make feasible such analysis, it was proposed in
Refs.~\cite{Mattos_2019,Mattos_2021,MATTOS} the modification in
the coupling constants of the model by making them functions of
$\Phi$ as
\begin{eqnarray}
G_s \rightarrow \mathcal{G}_s(G_s,\Phi) &=& G_s( 1-\Phi^2 ),
\\
G_V \rightarrow \mathcal{G}_V(G_V,\Phi) &=& G_V( 1-\Phi^2 ),
\\
K \rightarrow  \mathcal{K}(K,\Phi) &=& K( 1-\Phi^2 ),
\end{eqnarray}
where we have used the approximation
$\Phi=\Phi^*$~\cite{Mattos_2019,Mattos_2021,MATTOS}. The
motivation for these functions is to make vanishing all couplings
at the limit of the deconfined phase, attained at $\Phi\rightarrow
1$. By taking into account these assumptions we generate the
called PNJL0 model, for which pressure and energy density are
respectively given, at $T=0$, by
\begin{align}
&P_{\mbox{\tiny PNJL0}} = -G_s\sum _f \rho _{sf}^2 + G_V\sum _f \rho _f^2 - 4K\prod _f \rho _{sf} \nonumber\\
&+\frac{\gamma}{2\pi ^2}\sum _f\int_0^\Lambda
dk\,k^2(k^2+M_f^2)^{1/2} -\mathcal{U}(\rho_f,\rho_{sf},\Phi)
+\Omega_{\mbox{\tiny vac}}
\nonumber \\
&+\frac{\gamma}{6\pi ^2}\sum
_f\int_0^{k_{Ff}}\frac{dk\,k^4}{(k^2+M_f^2)^{1/2}}
\end{align}
and
\begin{align}
&\mathcal{E}_{\mbox{\tiny PNJL0}} = G_s\sum _f\rho _{sf}^2 + G_V
\sum _f\rho _{f}^2 + 4K\prod _f \rho _{sf}
\nonumber\\
&-\frac{\gamma}{2\pi^2} \sum _f\int_{k_{Ff}}^\Lambda
dk\,k^2(k^2+M_f^2)^{1/2} -2G_V\Phi^2\sum _f \rho _{f}^2
\nonumber\\
&  +\mathcal{U}(\rho_{(u,d,s)},\rho_{s(u,d,s)},\Phi) -
\Omega_{\mbox{\tiny vac}},
\end{align}
with the constant $\Omega_{\mbox{\tiny vac}}=-P_{\mbox{\tiny
vac}}$ is added to ensure $P_{\mbox{\tiny
PNJL0}}=\mathcal{E}_{\mbox{\tiny PNJL0}}=0$ in vacuum. In these
equations, one has quark masses and condensates written as
\begin{eqnarray}
M_f = m_f -2\mathcal{G}_s(G_s,\Phi)\rho
_{sf}-2\mathcal{K}(K,\Phi)\prod _{f'\neq f}\rho _{sf'},
\label{mfpnjl0}
\end{eqnarray}
and
\begin{eqnarray}
\rho_{sf}=-\frac{\gamma M_f}{2\pi^2}\int_{k_{Ff}}^\Lambda
\frac{dk\,k^2}{(k^2+M_f^2)^{1/2}}, \label{rhospnjl0}
\end{eqnarray}
with $\rho_f=(\gamma/6\pi^2)k_{Ff}^3$ and $\gamma=N_s\times N_c=6$
(degeneracy factor given in terms of spin, $N_s=2$, and color,
$N_c=3$, numbers). $\Lambda$ is the cutoff parameter and $m_f$ is
the current quark mass. The chemical potential of each quark reads
\begin{align}
\mu _f = (k_{Ff}^2 + M_f^2)^{1/2}+ 2\mathcal{G}_V(G_V,\Phi)\rho
_f.
\end{align}
Finally, the new Polyakov potential, that is not vanishing at
$T=0$, is defined here as
\begin{align}
&\mathcal{U}(\rho_{(u,d,s)},\rho_{s(u,d,s)},\Phi) = G_V\Phi^2\sum
_f \rho _f^2 -G_s\Phi^2\sum _f \rho _{sf}^2
\nonumber\\
&-4K\Phi^2\prod _f \rho _{sf} +
a_3T_0^4\mbox{ln}(1-6\Phi^2+8\Phi^3-3\Phi^4), \label{upnjl0}
\end{align}
for $T_0=190$~MeV. The last term in Eq.~\eqref{upnjl0} ensures
nonzero solutions for the traced Polyakov loop and also limits
this quantity to the range of $0<\Phi<1$. In order to determine
values for the quark condensates and for the traced Polyakov loop,
it is needed to solve, simultaneously, Eqs.~\eqref{mfpnjl0}
and~\eqref{rhospnjl0} in a self-consistent way, along with the
condition of
\begin{align}
\frac{\partial\Omega_{\mbox{\tiny PNJL0}}}{\partial\Phi}=0,
\end{align}
where $\Omega_{\mbox{\tiny PNJL0}}=-P_{\mbox{\tiny PNJL0}}$.

The constants used in this work for the PNJL0 model are given by
the Rehberg-Klevansky-Hufner parametrization, namely,
$G_s=3.67/\Lambda^2$, $K=-12.36/\Lambda^5$, $m_u=m_d=5.5$~MeV,
$m_s=140.7$~MeV, and $\Lambda=602.3$~MeV. The constants $G_V$ and
$a_3$ are the free parameters of the model. We verify in
Sec.~\ref{mesonprop} how the variation of these quantities affect
the meson properties.

\begin{figure}[!htb]
\centering
\includegraphics[scale=0.5]{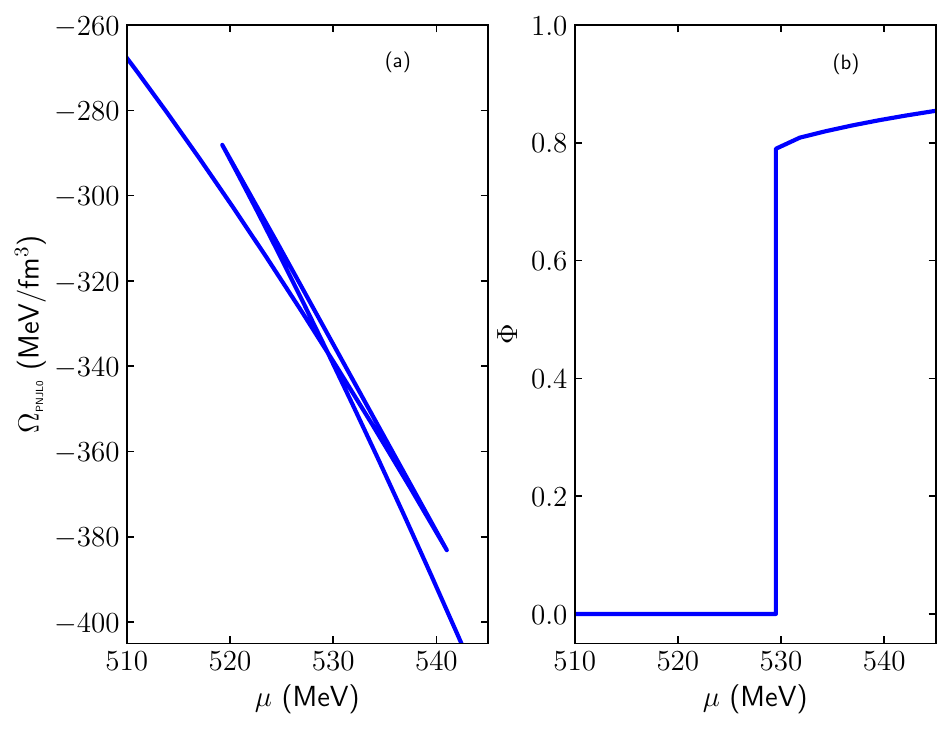}
\caption{(a) $\Omega_{\mbox{\tiny PNJL0}}$ and (b) $\Phi$ as a
function of the common chemical potential for the PNJL0 model with
$G_V/G_s=0.2$, and $a_3=-0.08$.} \label{ome-phi}
\end{figure}

\begin{figure}[!htb]
\centering
\includegraphics[scale=0.5]{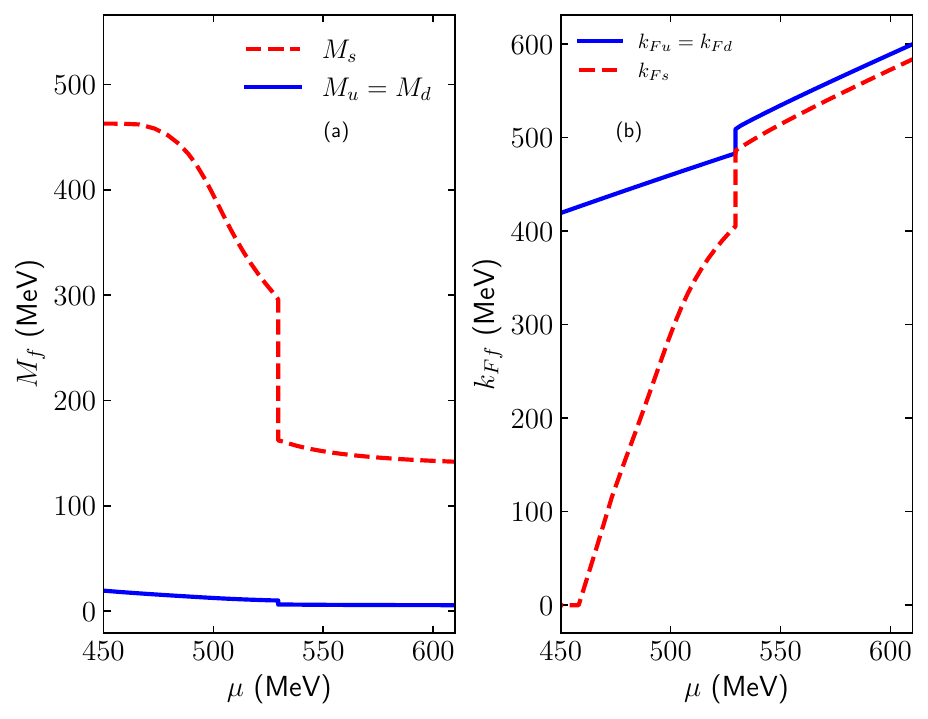}
\caption{(a) Constituent quark masses, and (b) quark Femi momenta
as a function of the common chemical potential for the PNJL0 model
with $G_V/G_s=0.2$, and $a_3=-0.08$.} \label{kF-M}
\end{figure}

For the calculation of the in medium single meson properties, we
focus in two particular systems constructed from the PNL0 model.
The first one consists of symmetric quark matter, for which the
chemical potentials are equal, i.e., we take
$\mu_u=\mu_d=\mu_s\equiv\mu$. The identification of the first
order confinement/deconfinement phase transition can be clearly
made by looking at the chemical potential dependence of the grand
canonical potential density, as shown in
Fig.~\ref{ome-phi}{\color{blue}a}.
This figure shows the signature of a first order phase transition,
for which the traced Polyakov loop is the respective order
parameter. As one can verify in Fig.~\ref{ome-phi}{\color{blue}b},
$\Phi$ abruptly increases from $\Phi=0$ to $\Phi>0$ at the
chemical potential related to the transition, in this case,
$\mu=529.5$~MeV. To determine the physical equation of state we
use a Maxwell construction connecting the stable regions at both
sides of the transition point. The appearing of nonvanishing
solutions of $\Phi$ leads to the effects depicted in
Fig.~\ref{kF-M} related to the constituent quark masses, and quark
Fermi momenta, respectively. It is clear that deconfinement leads
to a substantial decreasing of $M_s$, leading the system to the
direction of restoration of chiral symmetry. The Fermi momenta are
also modified by $\Phi$. Once again, the biggest change is
observed for the strange flavor.

The second case analyzed here is the charge neutral system of
quarks and leptons, namely, muons and massless electrons, in weak
equilibrium. For this system, the relationship between the
chemical potentials of quarks and leptons reads
\begin{eqnarray}
\mu_u &=& \frac{\mu_B}{3} - \frac{2}{3}\mu_e, \\
\mu_d &=& \mu_s = \frac{\mu_B}{3} + \frac{1}{3}\mu_e,\\
\mu_e &=& \mu_\mu
\end{eqnarray}
where the indexes $e$ and $\mu$ refer to electrons and muons,
respectively. The baryonic chemical potential is given by $\mu_B$.
The charge neutrality of the system leads to
\begin{eqnarray}
\frac{2}{3}\rho_u - \frac{1}{3}\rho_d - \frac{1}{3}\rho_s =
\frac{\mu_e^3}{3\pi^2} + \frac{(\mu_\mu^2 -
m_\mu^2)^{3/2}}{3\pi^2},
\end{eqnarray}
with $m_\mu=105.7$~MeV. For this system, we apply the same method
performed in Ref.~\cite{MATTOS}, namely, a hadron-quark phase
transition using the PNJL0 model for the quark sector. For the
hadronic side, we use a density-dependent model described the
Lagrangian density given by
\begin{align}
&\mathcal{L}_{\mbox{\tiny HAD}} =
\overline{\psi}(i\gamma^\mu\partial_\mu - M)\psi +
\Gamma_\sigma(\rho)\sigma\overline{\psi}\psi -
\Gamma_\omega(\rho)\overline{\psi}\gamma^\mu\omega_\mu\psi
\nonumber\\
&-\frac{\Gamma_\rho(\rho)}{2}\overline{\psi}\gamma^\mu\vec{\rho}_\mu\vec{\tau}\psi
+ \Gamma_\delta(\rho)\overline{\psi}\vec{\delta}\vec{\tau}\psi +
\frac{1}{2}(\partial^\mu \sigma \partial_\mu \sigma -
m^2_\sigma\sigma^2)
\nonumber\\
&- \frac{1}{4}F^{\mu\nu}F_{\mu\nu} +
\frac{1}{2}m^2_\omega\omega_\mu\omega^\mu
-\frac{1}{4}\vec{B}^{\mu\nu}\vec{B}_{\mu\nu}+\frac{1}{2}m^2_\rho
\vec{\rho}_\mu\vec{\rho}^\mu
\nonumber\\
&+ \frac{1}{2}(\partial^\mu\vec{\delta}\partial_\mu\vec{\delta} -
m^2_\delta\vec{\delta}^2), \label{dldd}
\end{align}
with
\begin{eqnarray}
\Gamma_i(\rho) =
\Gamma_i(\rho_0)a_i\frac{1+b_i(\rho/\rho_0+d_i)^2}{1+c_i(\rho/\rho_0+d_i)^2},
\label{gamadefault}
\end{eqnarray}
for $i=\sigma,\omega$, and
\begin{eqnarray}
\Gamma_i(\rho)=\Gamma_i(\rho_0)[a_ie^{-b_i(\rho/\rho_0-1)} -
c_i(\rho/\rho_0 - d_i)]. \label{gamarho}
\end{eqnarray}
for $i=\rho,\delta$. $\psi$ is the nucleon field and $\sigma$,
$\omega^\mu$, $\vec{\rho}_\mu$, and $\vec{\delta}$ are the scalar,
vector, isovector-vector, and isovector-scalar fields,
respectively. The antisymmetric tensors $F_{\mu\nu}$ and
$\vec{B}_{\mu\nu}$ are defined as
$F_{\mu\nu}=\partial_\nu\omega_\mu-\partial_\mu\omega_\nu$ and
$\vec{B}_{\mu\nu}=\partial_\nu\vec{\rho}_\mu-\partial_\mu\vec{\rho}_\nu$.
Nucleon rest mass, and mesons masses are given by $M_{\mbox{\tiny
nuc}}$, $m_\sigma$, $m_\omega$, $m_\rho$, and $m_\delta$.

The mean-field approximation is used once again in order to
compute energy density and pressure of the model, namely,
\begin{align}
\mathcal{E}_{\mbox{\tiny HAD}} &= \frac{1}{2}m^2_\sigma\sigma^2 -
\frac{1}{2}m^2_\omega\omega_0^2 -
\frac{1}{2}m^2_\rho\bar{\rho}_{0(3)}^2 +
\frac{1}{2}m^2_\delta\delta^2_{(3)}
\nonumber \\
&+\frac{\Gamma_\rho(\rho)}{2}\bar{\rho}_{0(3)}\rho_3
+\frac{1}{\pi^2}\int_0^{{k_F}_{p}}dk\,k^2(k^2+M_p^{*2})^{1/2}
\nonumber \\
&+\Gamma_\omega(\rho)\omega_0\rho +
\frac{1}{\pi^2}\int_0^{{k_F}_{n}}dk\,k^2(k^2+M_n^{*2})^{1/2}
\label{denergdd}
\end{align}
and
\begin{align}
P_{\mbox{\tiny HAD}} &= \rho\Sigma_R(\rho)-
\frac{1}{2}m^2_\sigma\sigma^2 + \frac{1}{2}m^2_\omega\omega_0^2 +
\frac{1}{2}m^2_\rho\bar{\rho}_{0(3)}^2
\nonumber \\
&- \frac{1}{2}m^2_\delta\delta^2_{(3)}
+\frac{1}{3\pi^2}\int_0^{{k_F}_{p}}\hspace{-0.5cm}\frac{dk\,k^4}{(k^2+M_p^{*2})^{1/2}}
\nonumber \\
&+\frac{1}{3\pi^2}\int_0^{{k_F}_{n}}\hspace{-0.5cm}\frac{dk\,k^4}{(k^2+M_n^{*2})^{1/2}}
\label{pressuredd}
\end{align}
with
\begin{align}
\Sigma_R(\rho)&=\frac{\partial\Gamma_\omega}{\partial\rho}\omega_0\rho
+\frac{1}{2}\frac{\partial\Gamma_\rho}{\partial\rho}\bar{\rho}_{0(3)}\rho_3
-\frac{\partial\Gamma_\sigma}{\partial\rho}\sigma\rho_s
\nonumber\\
&-\frac{\partial\Gamma_\delta}{\partial\rho}\delta_{(3)}\rho_{s3},
\end{align}
being the rearrangement term. The total scalar density is
\begin{align}
\rho_s &=  \rho_{sp} + \rho_{sn}
\nonumber\\
&=\frac{M^*}{\pi^2}\left[\int_0^{{k_F}_{p}}\hspace{-0.5cm}
\frac{dk\,k^2}{\sqrt{k^2+M_p^{*2}}}
+\int_0^{{k_F}_{n}}\hspace{-0.5cm}
\frac{dk\,k^2}{\sqrt{k^2+M_n^{*2}}}\right],
\end{align}
with $\rho_3=\rho_p-\rho_n$, and
$\rho_{p,n}={k_F}_{p,n}^3/3\pi^2$. The fields are obtained as
$\sigma = \Gamma_\sigma(\rho)\rho_s/m^2_\sigma$, $\omega_0 =
\Gamma_\omega(\rho)\rho/m_\omega^2$, $\bar{\rho}_{0(3)} =
\Gamma_\rho(\rho)\rho_3/2m_\rho^2$, and
$\delta_{(3)}=\Gamma_\delta(\rho)\rho_{s3}/m_\delta^2$ with
$\rho_{s3}=\rho_{sp}-\rho_{sn}$. Finally, the nucleon effective
masses are
\begin{eqnarray}
M^*_{p,n}=M_{\mbox{\tiny nuc}}-\Gamma_\sigma(\rho)\sigma\pm
\Gamma_\delta(\rho)\delta_{(3)}
\end{eqnarray}
with~$-(+)$ for protons (neutrons). The parametrization of the
density dependent model used here is the \mbox{DDH$\delta$}
one~\cite{gaitanos,debora2009,fortin}.

In the case of beta equilibrated matter, the conditions that need
to be satisfied in the hadronic side are the following:
$\rho_p-\rho_e=\rho_\mu$ and $\mu_n-\mu_p=\mu_e=\mu_\mu$. In this
case, the baryonic chemical potential is equal to the neutron
chemical potential ($\mu_B=\mu_n$).

\begin{figure}[!htb]
\centering
\includegraphics[scale=0.5]{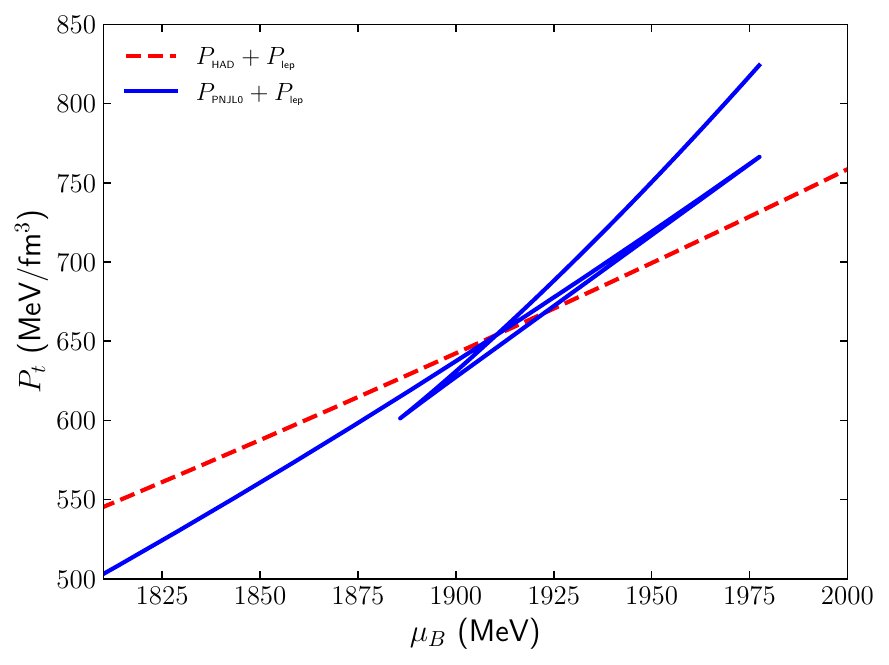}
\caption{Total pressure of beta equilibrated matter as a function
of $\mu_B$. For the PNJL0 model we take $G_V/G_s=0.5$, and
$a_3=-0.458$.} \label{press-mu}
\end{figure}

One way to perform the hadron-quark phase transition is through
the Maxwell construction between the \mbox{DDH$\delta$} model and
the PNJL0 one. In this case, pressure and chemical potential of
both phases are forced to be equal. The implementation of such an
approach is displayed in Fig.~\ref{press-mu}.
The total pressure is given by $P_t=P_{\mbox{\tiny
HAD}}+P_{\mbox{\tiny lep}}$ in the dashed curve, and by
$P_t=P_{\mbox{\tiny PNJL0}}+P_{\mbox{\tiny lep}}$ in the solid
one, with $P_{\mbox{\tiny lep}}$ being the sum of the electron and
muon pressures. As in Ref.~\cite{MATTOS}, we take the value of
$a_3$ in the PNJL0 model that makes the hadron-quark phase
transition coincide with the confinement/deconfinement one. The
full curve in this case is formed by the dashed one from
$\mu_B<\mu_{Bc}$, and by the solid one in the range of
$\mu_B>\mu_{Bc}$. One has $\mu_{Bc}=1910.4$~MeV for the parameter
set used to construct this figure.

\section{In medium properties of the pseudo-scalar mesons}
\label{mesonprop}

In this section we summarize the theoretical foundations of the
single meson properties evaluated in the next sections. We are
considering different frameworks, namely, within the PNJL0 model
the mesons emerge as a quasiparticle of the quark-antiquark
interaction. In the hadronic model, instead, the mesons are
preexisting degrees of freedom which are dressed by their
interaction with baryons. However, both situations are
non-perturbative and are solved by summing a given diagram to all
orders in a self-consistent approach. We consider here the
lightest pseudoscalar meson nonet at zero temperature and for a
density regime around the deconfinement phase transition.

\subsection{Quark model}

The mesonic modes have been studied in the SU(3) NJL model from
long time ago~\cite{KLEVANSKY} and developed
subsequently~\cite{REHBERG, COSTA} even in the extended PNJL model
at finite temperature~\cite{COSTA2}. For this purpose, the
specific channel responsible for the excitation of a given mesonic
mode is diagrammatically summed to all orders to build the
polarization insertion $\Pi(p)$ within a random phase
approximation (RPA).  To accomplish this strategy, the
contribution coming from the determinant in Eq.~\eqref{dlpnjl}
must be rendered into an effective four quarks coupling by
contracting $\bar{q}\, q$ pairs. In this approach the interaction
giving rise to the meson modes is
\begin{align}
&\sum_{a=0}^8 \left[L_a \left(\bar{q} \,\lambda^a q\right)^2 + K_a
\left(\bar{q} \,i\gamma_5 \lambda^a q\right)^2\right]
\nonumber\\
&+\sum_{a,b=0,3,8} (1-\delta_{a,b}) \left[L_{a b}\; \bar{q}
\,\lambda^a q \;\bar{q} \lambda^b q \right.
\nonumber\\
&+\left. K_{a b}\; \bar{q} \,i\gamma_5 \lambda^a q \; \bar{q}
\,i\gamma_5 \lambda^b q\right].
\end{align}
The effective coupling constants are given by
\begin{eqnarray}
K_0={\cal G}_s/2-\left(\rho_{s u}+\rho_{s d}+\rho_{s s}\right)
{\cal K}/3,
\\
K_1=K_2=K_3=\left({\cal G}_s-\rho_{s s}{\cal K}\right)/2,
\\
K_4=K_5=\left({\cal G}_s-\rho_{s d}{\cal K}\right)/2,
\\
K_6=K_7=\left({\cal G}_s-\rho_{s u}{\cal K}\right)/2,
\\
K_8={\cal G}_s/2-\left(2 \rho_{s u}+2 \rho_{s d}-\rho_{s s}\right)
{\cal K}/6,
\\
K_{0 3}=K_{3 0}=\left(\rho_{s d}-\rho_{s u}\right) {\cal
K}/\sqrt{24},
\\
K_{0 8}=K_{8 0}= \left(\rho_{s u}+\rho_{s d}-2 \rho_{s s}\right)
{\cal K}/\sqrt{72},
\\
K_{3 8}=K_{8 3}=\left(\rho_{s u}-\rho_{s d}\right) {\cal
K}/\sqrt{12}.
\end{eqnarray}
The remaining ones $L_a,\, L_{a b}$ participate only of the scalar
meson construction and will not be used in the present
calculations. It must be noted that for $\rho_{s u}=\rho_{s d}$ we
have $K_{3 8}=K_{3 0}=0$ and the sector $3$ decouples form the
$0-8$, that is the neutral pion becomes independent of the
$\eta-\eta'$ mesons. Since the sector $0-3-8$ is generally
coupled, the calculations are performed in two different
instances. The charged pions and kaons properties are evaluated
independently, while the $\pi^0, \eta,$ and $\eta'$ are solved
together.

The polarization insertion connecting states $a$ and $b$ is given
by
\begin{align}
&i \Pi_{a b}(p)=
\nonumber\\
&=N_c  \sum_{j, k}\lambda_{j k}^a \lambda_{k j}^b \int
\frac{d^4q}{(2 \pi)^4} \text{Tr} \left\{ i\, \gamma_5 G_j(q)\,i\,
\gamma_5 G_k(p-q)\right\} \label{NJL POL}
\end{align}
which can be projected onto physical meson states as
\begin{align}
&i \Pi_{a}(p) =
\nonumber\\
&= N_c  \sum_{j, k} T_{j k}^a T_{k j}^a \int \frac{d^4q}{(2
\pi)^4} \text{Tr} \left\{ i\, \gamma_5 G_j(q)\,i\, \gamma_5
G_k(p-q)\right\}
\end{align}
where $T^a=\left(\lambda_1\pm i\,\lambda_2\right)/\sqrt{2}$ for
$a=\pi^\pm$, $T^a=\left(\lambda_6\pm i\,\lambda_7\right)/\sqrt{2}$
 for $a=K^0$ (upper sign) or $a=\bar{K}^0$ (lower sign), and
finally $T^a=\left(\lambda_4\pm i\,\lambda_5\right)/\sqrt{2}$ for
$a=K^\pm$.

The propagator $G_j(p)$ of the quark of flavor $j$ in a real time
formulation of the thermal field theory is written in the mean
field approach as
\begin{align}
&G_j(p)=
\nonumber\\
&=\left(\not\!\tilde{p}+M_j\right)\left[\frac{1}{\tilde{p}^2-M_j^2+i
\varepsilon} +2\,i\, \pi\, \Theta(\mu_j-\tilde{p}_0)\,
n_F(\tilde{p}_0)\right]\label{QProp}
\end{align}
where $\tilde{p}_\nu=(p_0-2\,{\cal G}_V \,\rho_j,\bm{p})$ and
$n_F$ is the Fermi occupation number. We use the same cutoff
$\Lambda$ to deal with the divergent contributions coming from the
Dirac sea. The explicit result for the integral in Eq. (\ref{NJL
POL}) is given in the case $\bm{p}=0$ by
\begin{widetext}
\begin{eqnarray}
&&\int \frac{d^4q}{(2 \pi)^4} \text{Tr} \left\{ i\, \gamma_5
G_j(q)\,i\, \gamma_5 G_l(p-q)\right\}=-\frac{\rho_{s
j}}{M_j}-\frac{\rho_{s l}}{M_l}+\frac{N_c}{2
\pi^2}\frac{\tilde{p}_0^2-(M_j-M_l)^2}{\tilde{p}_0} \nonumber \\
&&\times\Bigg[k_{F l}-k_{F j}+C_j\,\ln\left(\frac{k_{F j}+E_{F
j}}{\Lambda+E_{\Lambda j}}\right)-C_l\,\ln\left(\frac{k_{F l}+E_{F
l}}{\Lambda+E_{\Lambda l}}\right)+W\, \left(F_j-F_l\right)\Bigg]
\end{eqnarray}
where
\begin{eqnarray}
F_i=\left\{ \begin{array}{ll}\arctan(k_{F
i}/W)-\arctan(x_i)+\arctan(y_i),&C_i^2-M_i^2<0\\
\frac{1}{2}\ln\left|\frac{1-x_i}{1+x_i}\,\frac{1-k_{F i}/W}{1+k_{F
i}/W}\right|-\ln\left|\frac{1-y_i}{1+y_i}\right|,& C_i^2-M_i^2>0
\end{array}\right. \nonumber
\end{eqnarray}
\end{widetext}
and $E_{F i}=\sqrt{k_{F i}^2+M_i^2}$, $E_{\Lambda
i}=\sqrt{\Lambda^2+M_i^2}$,
$C_j=\left(M_l^2-M_j^2-\tilde{p}_0^2\right)/2 \tilde{p}_0$,
$C_l=C_j+\tilde{p}_0$,
$W=\sqrt{|C_j^2-M_j^2|}=\sqrt{|C_l^2-M_l^2|}$, $x_i=C_i k_{F
i}/W\,E_{F i}$, and $y_i=C_i \Lambda/W\,E_{\Lambda i}$. The meson
polarizations can have an imaginary part under certain conditions
which will be discussed in Sec.~\ref{RESULTS}, the equation above
shows only the real part.

The in medium meson mass $m_a$ is obtained as the solution of the
equation~\cite{REHBERG}
\begin{align}
0=1-4 K_a \Pi_a(p_0=m_a,\bm{p}=0), \label{zeromass}
\end{align}
in which the coupling must be chosen as $K_a=K_1$ for $\pi^\pm$,
$K_a=K_4$ for $K^\pm$, and $K_a=K_6$ for the neutral kaons.
Furthermore, the effective meson-quark-antiquark couplings
$G_{\alpha q}$ can be identified from a pole approximation as
\begin{align}
G_{\alpha q}^{-2}=\frac{1}{m}\; \frac{\partial
\Pi_{\alpha}}{\partial p_0}\Bigg]_{p_0=m}. \label{coupling}
\end{align}
In the case of the coupled $\pi^0 \eta \eta'$ sector, the
calculations must be done in a matrix framework by introducing the
quantities
\begin{eqnarray}
\mathbb{K}=\left(\begin{array}{ccc} K_3&K_{3 0}&K_{3 8}\\
K_{0 3}&K_0&K_{0 8}\\K_{8 3}&K_{8 0}&K_8\\
\end{array}\right),\; \mathbb{P}=\left(\begin{array}{ccc} \Pi_{3 3}&\Pi_{3 0}&\Pi_{3 8}\\
\Pi_{0 3}&\Pi_{0 0}&\Pi_{0 8}\\\Pi_{8 3}&\Pi_{8 0}&\Pi_{8 8}\\
\end{array}\right)
\end{eqnarray} and
$\mathbb{M}^{-1}=\mathbb{K}^{-1}-\mathbb{P}$. The last matrix has
a set of eigenvalues $\nu_i(p),\; i=1-3$ which are used to define
the mass $m$ through the solutions of $0=\nu_i(p_0=m,\bm{p}=0)$
for the $\pi^0$ ($i=1$), $\eta$ ($i=2$), and $\eta'$ ($i=3$).

\subsection{Hadronic model}

The study of the meson properties in a dense hadronic environment
is a longstanding issue~\cite{hayano}. The importance of the
variation of the constitutive properties of the lightest mesons
has been emphasized in different situations.  In nuclear physics
the role of the pion~\cite{BROWN_RHO2}, correlated pion
exchange~\cite{RAPP} and $\sigma$ meson~\cite{BROWN_RHO}, have
been remarked.  For the description of medium and large matter
densities the use of scalar and vector fields $\sigma,\, \omega,\,
\bar{\rho}, \, \delta$ has became a standard
procedure~\cite{SAITO}. The possibility of meson condensation in
neutron stars has intensified the study of the kaon
properties~\cite{GLENDENNING,TOLOS,KAPLAN, BROWN_LEE}, while the
analysis of exotic nuclei has regarded the interaction between
nucleons and $\pi, K,$ and $\eta$ mesons~\cite{FRIEDMAN,CIEPLY}.

A large number of such investigations are based on the
SU(3)$\times$SU(3) chiral Lagrangian, keeping a low order
expansion in inverse powers of the chiral symmetry breaking scale.
As a consequence the range of applicability is expected to extend
over a density range of several times the normal nuclear density
$\rho_0=0.16 \text{fm}^{-3}$. In addition, there are significant
research using effective models founded on semi-phenomelogical
considerations.

In a preceding publication~\cite{MATTOS}, a composition of the
PNJL0 model and the hadronic model with density dependent
couplings \mbox{DDH$\delta$} has been done in order to match the
threshold of the deconfinement transition. With the aim to
continue this approach we extend here the \mbox{DDH$\delta$} model
by including additional terms which consider the role of $\pi, \,
K$, and $\eta$ mesons. In the absence of pseudoscalar meson
condensation, these new contributions do not modify the Euler
equations used to obtain the results shown in Section \ref{form}.
Thus, we add to ${\cal L}_{\mbox{\tiny HAD}}$ the sum of three
independent contributions, namely
\begin{equation}
\mathcal{L}={\cal L}_\pi + {\cal L}_K + {\cal L}_\eta, \label{Add
Lagr}
\end{equation}
which have been selected for their ability to describe the main
features of the mesons in a dense medium in a concise formulation.

The first term is the Weinberg model of the pion-nucleon
interaction~\cite{WEINBERG}, proposed for low energy pion-nucleon
scattering and satisfies the Goldberger-Treiman relation. It reads
as
\begin{equation}{\cal L}_\pi=\frac{g_A}{2 f_\pi}\bar{\Psi}\,\gamma^\mu \gamma^5 \bm{\tau} \cdot \partial_\mu
\bm{\pi} \Psi-\frac{1}{ f_\pi^ 2}\bar{\Psi}\bm{\tau}\Psi \cdot
(\bm{\pi}\times \partial_\mu
\bm{\pi})\label{SCHWINGER}\end{equation}
The bi-spinor $\Psi$ has the proton and neutron components
$\Psi_1=\psi_p,\, \Psi_2=\psi_n$. The second term in the equation
above is known as the Weinberg-Tomozawa interaction and
contributes at lowest order only to the charged pion polarization.
The first term contributes at the second order with a bubble
diagram included in a RPA. Thus, one obtains
\begin{align}
&i \Pi_{\alpha}(p)=i \Pi_{\alpha}^{\text{OPE}}(p)+i
\Pi_{\alpha}^{\text{WT}}(p)
\\
&i\Pi_{\alpha}^{\text{OPE}}(p)=\left(\frac{g_A}{2 f_\pi}\right)^2
p_\mu p_\nu\times
\nonumber\\
&\times\sum_{a,b=1,2}\,\xi_\alpha\int\,\frac{d^4q}{(2 \pi)^4}\,
\text{Tr}\left[ \gamma^\mu \gamma_5 G^{(a)}(q)\, \gamma^ \nu
\gamma_5 G^{(b)}(q-p) \right]
\\%
&i\Pi_{\alpha}^{\text{WT}}(p)=-\frac{\epsilon_{\alpha}}{f_\pi^2}\,
p_\mu\,\sum_{c=1,2} \tau_3^{cc} \int\,\frac{d^4q}{(2 \pi)^4}\,
\text{Tr}\left[ \gamma^ \mu G^{(c)}(q)  \right].
\end{align}
The index $\alpha$ discriminates the isotopic component, for the
neutral pion is $\epsilon_\alpha=0, \; \xi_\alpha=\delta_{a,b}$,
for the $\pi^-$ one must take $\epsilon_\alpha=1, \;
\xi_\alpha=2\,\delta_{a,2}\delta_{b,1}$ and finally for the
$\pi^+$ is $\epsilon_\alpha=-1, \;
\xi_\alpha=2\,\delta_{a,1}\delta_{b,2}$. The nucleon propagators
have, in a real time formulation of the thermal field theory, an
expression similar to Eq. (\ref{QProp})
\begin{align}
&G^{(a)}(p)=
\nonumber\\
&=\left(\not\!\tilde{p}+M_a^*\right)\left[\frac{1}{\tilde{p}^2-M_a^{*\,
2}+i \varepsilon} +2\,i\, \pi\, \Theta(\mu_a-\tilde{p}_0)\,
n_F(\tilde{p}_0)\right]
\end{align}
but with
\begin{align}
\tilde{p}_\mu=\left(p_0-\Gamma_\omega \omega_0\mp
\frac{\Gamma_\rho}{2}\,\rho_{0 (3)}-\Sigma_R,\bm{p}\right).
\end{align}
The explicit formulae for these components are easily obtained as
\begin{widetext}
\begin{eqnarray}
\Pi_{\alpha}^{\text{OPE}}(p_0,\bm{p}=0)&=&\left(\frac{g_A p_0}{2
f_\pi \tilde{p}_0}\right)^2
 \sum_{a,b=1,2}\,\xi_\alpha \Bigg\{\frac{1}{2}\left(M_a^{*\, 2}-M_b^{* \,
 2}\right)\left(\frac{\rho_{s b}}{M_a^*}-\frac{\rho_{s a}}{M_b^*}\right)+\tilde{p}_0 \left(\rho_{b}-\rho_{a}\right)
 \nonumber \\
&+&\left(\frac{M_a^*+M_b^*}{2
\pi}\right)^2\frac{\left(M_a^*-M_b^*\right)^2-\tilde{p}_0^2}{\tilde{p}_0}\,
\Bigg[k_{F b}-k_{F a}+C_a \ln\left(\frac{k_{F a}+E_{F
a}}{M_a^*}\right)
\nonumber\\
&-&C_b \ln\left(\frac{k_{F b}+E_{F b}}{M_b^*}\right)+W
\left(F_b-F_a\right)\Bigg] \Bigg\},
\\
\Pi_{\alpha}^{\text{WT}}(p_0,\bm{p}=0)&=&\frac{\epsilon_\alpha}{2
f_\pi^2}\left(\rho_p-\rho_n\right).
\end{eqnarray}
\end{widetext}%
Here only the real part of $\Pi^{\text{OPE}}$ is shown, while the
effects of the imaginary term are discussed in section
\ref{RESULTS}. As usually done in this framework, the divergent
contribution coming from the Dirac sea has been neglected.\\
The following definitions have been used
\begin{eqnarray}
F_c=\left\{ \begin{array}{ll}\-\arctan(k_{F
c}/W)+\arctan(x_c),&C_k^2-M_k^{*\, 2}<0\\
\frac{1}{2}\ln\left|\frac{1+x_c}{1-x_c}\,\frac{1-k_{F c}/W}{1+k_{F
c}/W}\right|, & C_k^2-M_k^{*\, 2}>0 \end{array}\right. ,\nonumber
\end{eqnarray}
together with $C_b=\left(M_a^{*\, 2}-M_b^{* \,
2}+\tilde{p}_0^2\right)/2 \tilde{p}_0$, $C_a=C_b-\tilde{p}_0$,
$W=\sqrt{|C_b^2-M_b^{* \, 2}|}$, $E_{F c}=\sqrt{k_{F c}^2+M_c^{*
\, 2}}$, and $x_c=C_c k_{F c}/W\,E_{F c}$.

The second term in Eq. (\ref{Add Lagr}) corresponds to the
kaon-nucleon interaction and is taken from~\cite{MISHRA_SANYAL}.
The model is based on a SU(3) chiral symmetric Lagrangian plus
symmetry breaking terms respecting the partial conservation of the
axial current, taken in the MFA~\cite{PAPAZOGLOU}. The relevant
terms for our calculations are
\begin{widetext}
\begin{eqnarray}
{\cal L}_K&=&\frac{-i}{4 f_K^2}\Big[\left(2 \bar{\psi}_1
\gamma^\mu \psi_1+ \bar{\psi}_2\gamma^\mu
\psi_2\right)\left(K^-\partial_\mu K^+ -K^+\partial_\mu K^
-\right) +\left(\bar{\psi}_1 \gamma^\mu \psi_1+2
\bar{\psi}_2\gamma^\mu \psi_2\right)
\left(\bar{K}^0\partial_\mu K^0-K^0 \partial_\mu \bar{K}^ 0\right)\Big]\nonumber \\
&+&\frac{m_K^2}{2 f_K}\left[ (\sigma+\delta)\,K^+
K^-+(\sigma-\delta)\,K^0
\bar{K}^0\right]-\frac{1}{f_K}\left[(\sigma+\delta)\,\partial_\mu
K^+ \partial^\mu K^-
+(\sigma-\delta)\,\partial_\mu K^0 \partial^\mu \bar{K}^0\right]\nonumber\\
&+&\frac{d_1}{2 f_K^2}(\bar{\psi}_1  \psi_1+2 \bar{\psi}_2
\psi_2)\,\partial_\mu K^0
\partial^\mu \bar{K}^0+\frac{d_2}{2 f_K^2}\left(\bar{\psi}_1  \psi_1
\,\partial_\mu K^+ \partial^\mu K^-+ \bar{\psi}_2 \psi_2
\,\partial_\mu K^0 \partial^\mu \bar{K}^0\right).
\end{eqnarray}
\end{widetext}
Treating the first order contribution to the kaon polarization,
the following results are obtained~\cite{MISHRA_SANYAL}
\begin{align}
 &\Pi_{K^\pm}(p)=\mp\frac{p_0}{2 f_K^2}(2 \rho_p+\rho_n)+\frac{m_K^2}{2 f_K}\,
 (\sigma+\delta)+(p_0^2-\bm{p}^2)\,\times
 \nonumber\\
 &\times\left[-\frac{\sigma+\delta}{f_K}+\frac{d_1}{2 f_K^2}\,\left(\rho_{s
p}+\rho_{s n}\right)+\frac{d_2}{2 f_K^2}\,\rho_{s p}\right],
\end{align}
and
\begin{align}
&\Pi_{K^0}(p)=\mp\frac{p_0}{2 f_K^2}(\rho_p+2
\rho_n)+\frac{m_K^2}{2
f_K}\,(\sigma-\delta)+(p_0^2-\bm{p}^2)\times
\nonumber\\
&\times\left[-\frac{\sigma-\delta}{f_K}+\frac{d_1}{2
f_K^2}\,\left(\rho_{s p}+\rho_{s n}\right)+\frac{d_2}{2
f_K^2}\,\rho_{s n}\right].
\end{align}
In the last equation the upper (lower) sign corresponds to the
$K^0$ ($\bar{K}^0$).

In order to match the prescriptions given in~\cite{MISHRA_SANYAL},
we have identified the fluctuations around the vacuum values of
the $\sigma$ and $\delta$ fields with the mean field values
described in Sec.~\ref{form}. The numerical values of the
parameters are $f_K=122$ MeV, $d_1=2.5/m_K$, and $d_2=0.72/m_K$.

The last term of Eq. (\ref{Add Lagr}) stands for the nucleon-eta
coupling, which we chose as proposed in~\cite{ZHONG}. It results
from a low momentum approach to the s-wave $\eta N$ scattering
starting from the chiral Lagrangian, and is given by
\begin{align}
{\cal L}_\eta=\frac{1}{2 f_\pi^2}\,\bar{\Psi}\left(\Sigma_{\eta N}
\;\eta^2+\kappa\; \partial_\mu \eta \, \partial^\mu
\eta\right)\Psi.
\end{align}
The fact that this interaction is quadratic in the meson field can
be used, as in the kaon case, to correct its propagator with a
first order diagram. The polarization insertion obtained in such
approach is
\begin{align}
\Pi_\eta(p)=-\frac{1}{f_\pi^2}\,\left(\Sigma_{\eta
N}+\kappa\,p_\mu p^\mu\right)\, \left(\rho_{s p}+\rho_{s
n}\right),
\end{align}
and the model parameters are fixed as $\Sigma_{\eta N}=283$ MeV
and $\kappa=0.4$. For each meson described by the interactions of
Eq. (\ref{Add Lagr}), the in-medium effective mass $m^*_X$ of the
$X=\pi,\,K,\,\eta$ meson is determined by solving the equation
\begin{align}
0=m^{*\, 2}_X-m_X^2-\Pi_X(p_0=m_X^*,\bm{p}=0),
\end{align}
where $m_X$ represents the in-vacuum mass.

\section{Results and discussion}
\label{RESULTS}

The total baryon number is a conserved charge of the strong
interaction, therefore if $\rho_B$  is used for the baryon number
per unit volume and $\rho_q$ for the particle number density for
each flavor, then the local constraint $\rho_B=\sum_q \rho_q/3$
always holds in quark matter. In the following we show and discuss
the results for the meson properties under two different flavor
compositions of homogeneous dense matter. In the first case we
consider independent flavor conservation, so that a chemical
potential $\mu_q$ is assigned to each flavor present in the Fermi
sea, and the complementary condition $\mu_u=\mu_d=\mu_s$ is
imposed. This configuration is named as symmetric flavor matter
(SFM), although it does not imply equal proportions of all the
flavors. In particular, the strange quark could be absent if $M_s>
\mu_{u,\,d}$, circumstantially it would present a threshold point
where this inequality is reversed. In the second case we focus on
matter under local electric neutrality, with the possible
contribution of electrons, a situation denoted in the following as
stellar matter (SM).

The PNJL0 and NJL results coincide in the low density regime,
where $\Phi=0$ and there exist a well known instability for quark
matter. It has been identified as a partial restoration of the
chiral symmetry, which for low values of the parameter $G_V/G_s$
gives rise to a first order phase transition. As $G_V/G_s$ is
raised, the transition becomes a smooth crossover. The behavior of
the pseudoscalar mesons within this domain has been studied for
instance in \cite{ABUKI,COSTA,COSTA3}, therefore, we do not
consider here its explicit analysis.

\begin{figure}[!htb]
\centering
\includegraphics[scale=0.5]{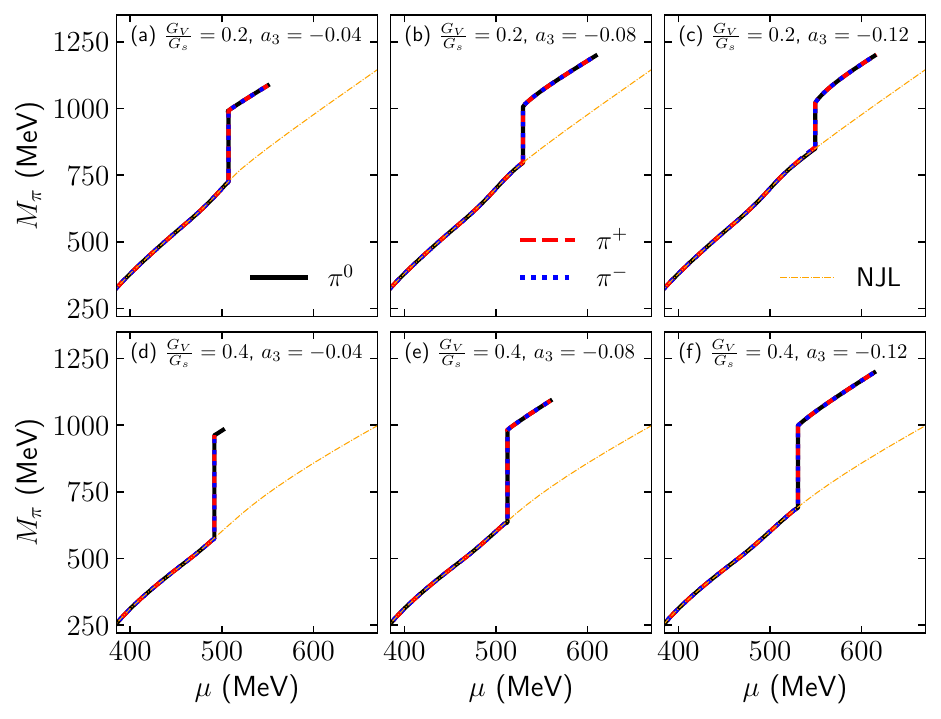}
\includegraphics[scale=0.5]{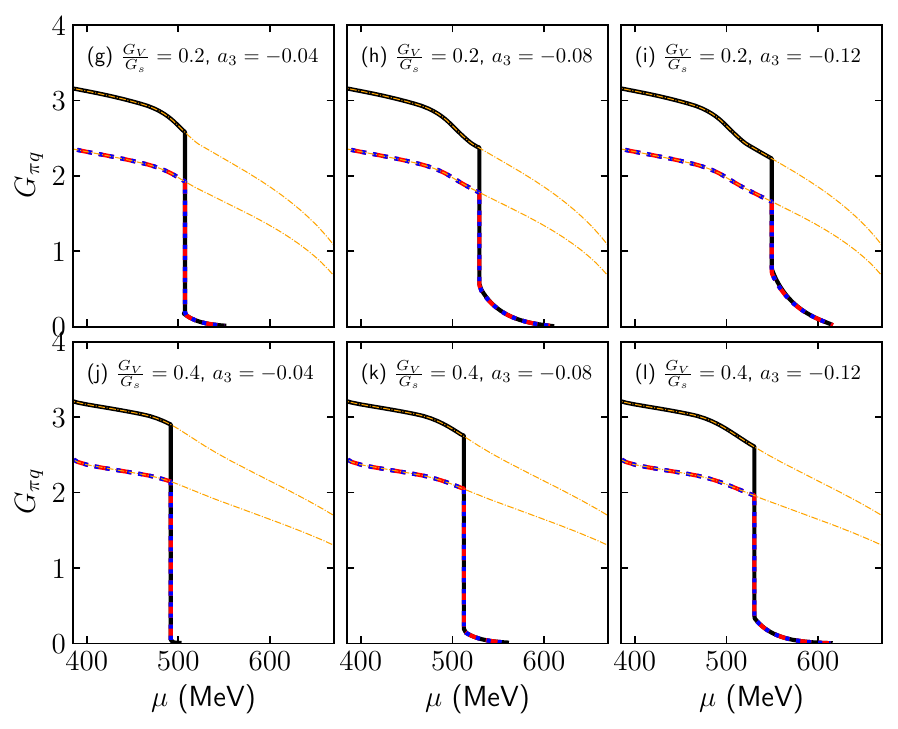}
\caption{Pion masses, and pion-quark couplings obtained from the
PNJL0 model (symmetric matter case) for different
parametrizations, namely, ($G_V/G_s$, $a_3$) given by (a,g)~$(0.2,
-0.04)$, (b,h)~$(0.2, -0.08)$, (c,i)~$(0.2,-0.12)$,
(d,j)~$(0.4,-0.04)$, (e,k)~$(0.4,-0.08)$, and (f,l)~$(0.4,-0.12)$.
Dotted-dashed lines: NJL model results.} \label{sym-pions}
\end{figure}

We start this discussion considering SFM in the PNJL0 model. The
results for the pions are displayed in Fig.~\ref{sym-pions}.
It is possible to see gaps in $M_\pi$ and $G_{\pi q}$ due to the
transition to the deconfined phase that takes place in the PNJL0
model. This feature is present for different values of $G_V$ and
$a_3$, taken as free parameters of the model. In the case of the
pion masses, the appearance of nonzero solutions of the traced
Polyakov loop induces a significant increase in $M_\pi$. The
opposite happens for the pion-quark couplings, namely, an abrupt
decrease occurs at the chemical potential in which the transition
occurs. Because of the degeneracy in the light $u-d$ sector, all
curves corresponding to the different isotopic components
coalesce. An almost linear increase with the chemical potential is
obtained for all the cases, and particularly at the deconfinement
threshold such increase takes place discontinuously. The gap in
the pion mass increase with both $G_V$ and $a_3$, and varies from
$200$ MeV to $380$ MeV. At the transition point the pion is not
longer a light excitation, since it is between 4-6 times heavier
than at zero density. A measure of the interaction of the pion
with its environment is given by $G_{\pi q}$, its effective
coupling with the quark-antiquark pair, shown in the bottom panels
of Fig.~\ref{sym-pions}. It can be appreciated that the greater
value of $G_V$ stabilize the density dependence of the remnant
interaction of the pions, at both sides of the transition. A
drastic reduction of $G_{\pi q}$ at the transition point leads to
a regime of almost non-interacting pions. An increase of $|a_3|$
attenuates this drop, as is more evident for the $G_V/G_s=0.2$
case.

\begin{figure}[!htb]
\centering
\includegraphics[scale=0.5]{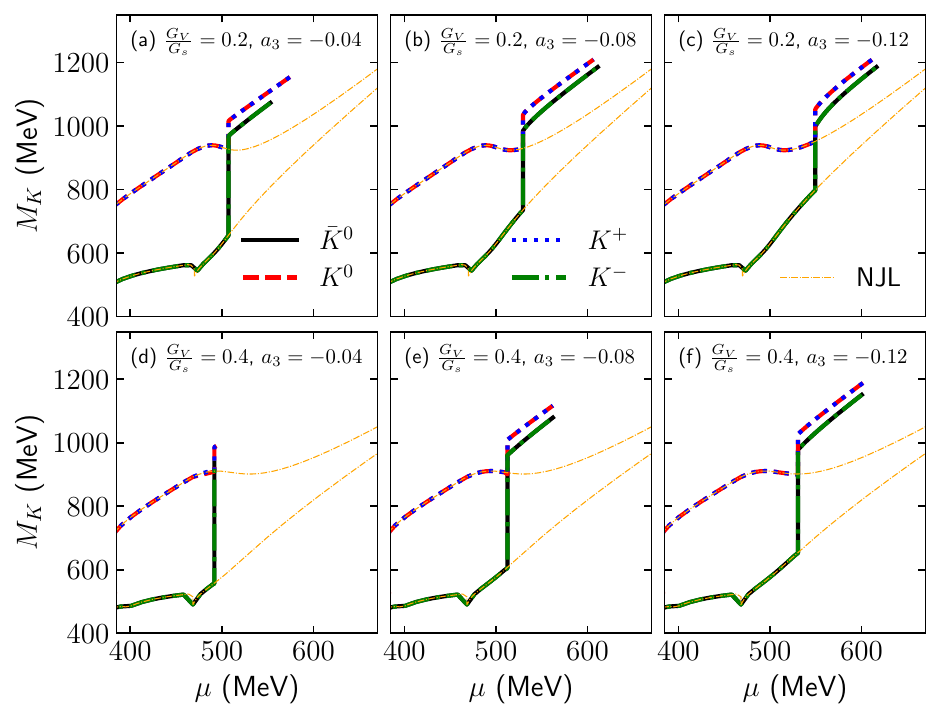}
\includegraphics[scale=0.5]{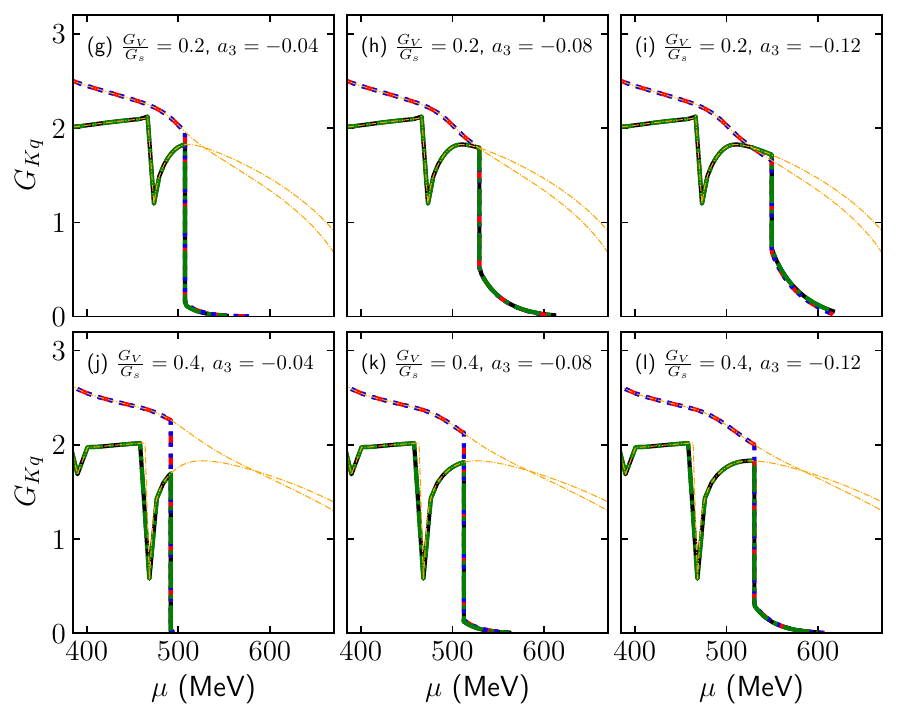}
\caption{Kaon masses, and kaon-quark couplings obtained from the
PNJL0 model (symmetric matter case) for different
parametrizations, namely, ($G_V/G_s$, $a_3$) given by (a,g)~$(0.2,
-0.04)$, (b,h)~$(0.2, -0.08)$, (c,i)~$(0.2,-0.12)$,
(d,j)~$(0.4,-0.04)$, (e,k)~$(0.4,-0.08)$, and (f,l)~$(0.4,-0.12)$.
Orange dotted-dashed lines: NJL model results.} \label{sym-kaons}
\end{figure}

The variation of the kaon mass in the SFM is displayed in Fig.
~\ref{sym-kaons}.
The curves corresponding to $\bar{K}^0$ and $K^-$ coincide as well
as those for $K^0$ and $K^+$ do, because of the symmetry of the
non-strange quarks. Within the pure NJL model the onset of the
strange quark manifests by a slowing down of the increase of the
masses $M_{K_0}$ and $M_{K^+}$, in contrast with the quick
variation of $M_{\bar{K}^0}$ and $M_{K^-}$ with the chemical
potential. The difference can be attributed to the fact that in
the former case the strange quark enters as an antiquark and does
not follow directly the sudden changes of the valence quarks. The
strangeness onset takes place around $\mu \simeq 460$ MeV, which
corresponds to a baryonic density $\rho_B/\rho_0\simeq 4.3$ for
$G_V/G_s=0.2$ and $\rho_B/\rho_0\simeq 3.7$ for $G_V/G_s=0.4$.
This feature contrasts with other calculations where the role of
the valence strange quarks are neglected ~\cite{BERNARD,COSTA}.
The non-monotonous dependence of the masses of the kaons
$K^0,\,K^+$ extends to higher densities, and part of such behavior
takes place within the instability region. Hence, it is partially
suppressed by the Maxwell construction performed within the PNJL0.
Since an increase of $|a_3|$ shifts the transition point towards
higher densities, the cases with $a_3=-0.12$ preserve more
complete this pattern, caused by the strange quark in the NJL
model. At the transition threshold a discontinuous increase of the
masses of all the kaons happens in a different manner. The mass of
the mesons with strangeness $S=1$ experience a gap around $100$
MeV, while the change is more abrupt for the $S=-1$ case, with a
gap taking values from $200$~MeV to $400$~MeV approximately. This
result is valid for all the parametrizations studied. Beyond that
point, a linear increase follows with very similar values for all
the isotopes and the masses are greater than $1$~GeV.

The effective coupling of the kaon with a $q\,\bar{q}$ pair is
shown in the lower panels of Fig.~\ref{sym-kaons}. As in the case
of pions, one can distinguish the evolution from a phase of mesons
interacting with its environment to a gas of quasi-free bosons.
The activation of the valence strange quarks manifests by a sudden
decrease of the effective coupling of the $S=-1$ kaons, followed
by a similarly quick restoration of its previous value. This
critical variation of $G_{K q}$ has been found for instance in
\cite{BERNARD}, although in our calculations there is no a
complete collapse but we have found multiple solutions at that
point. With regard to the local minimum shown by the effective
coupling in Figs.~\ref{sym-kaons}{\color{blue}j},
\ref{sym-kaons}{\color{blue}k}, and~\ref{sym-kaons}{\color{blue}l}
at $\mu$ slightly below $400$~MeV, it can be identified as a
threshold effect due to the fact that the meson excitation becomes
a resonance ~\cite{SOUSA}.

\begin{figure}[!htb]
\centering
\includegraphics[scale=0.5]{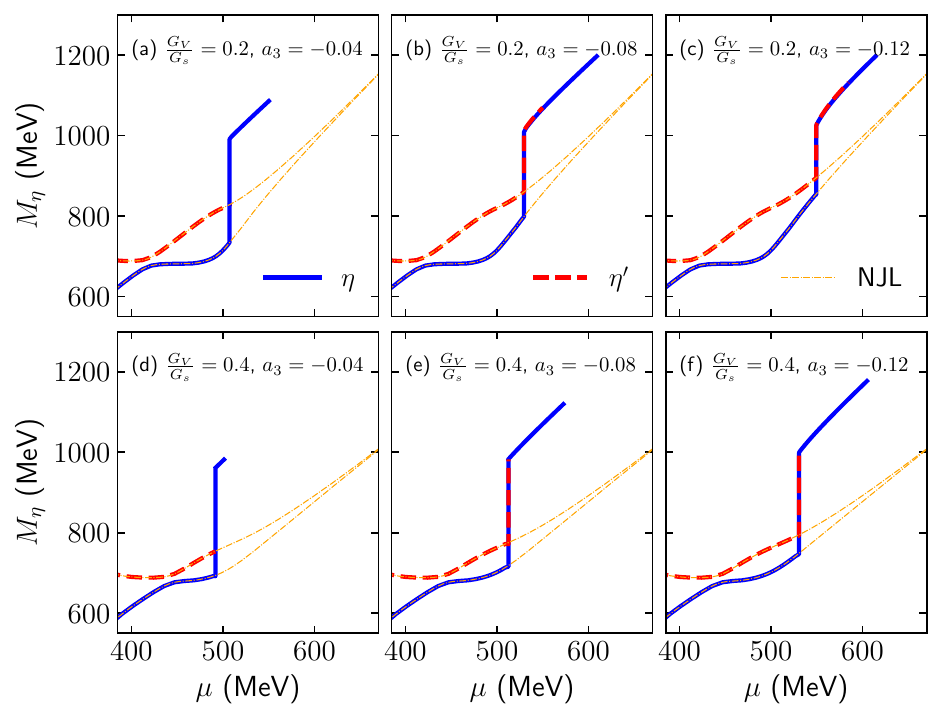}
\caption{$\eta$ and $\eta'$ masses obtained from the PNJL0 model
(symmetric matter case) for different parametrizations, namely,
($G_V/G_s$, $a_3$) given by (a)~$(0.2, -0.04)$, (b)~$(0.2,
-0.08)$, (c)~$(0.2,-0.12)$, (d)~$(0.4,-0.04)$, (e)~$(0.4,-0.08)$,
and (f)~$(0.4,-0.12)$. Orange dotted-dashed lines: NJL model
results.} \label{sym-etas_mass}
\end{figure}

The masses of the $\eta- \eta'$ system shown in
Fig.~\ref{sym-etas_mass} exhibits a similar variation as described
for pions and kaons. The coincidence with the NJL results are
extended to higher densities as $|a_3|$ grows, and a discontinuous
increment is experienced at the transition point followed by a
linear increase with degenerate masses.

It must be pointed out that within our approach and using the
specific parametrization shown in
Fig.~\ref{sym-etas_mass}{\color{blue}d}, all the mesonic
excitations are extinguished in the deconfined phase, since the
solutions for the effective masses cease to exist.

The discussion of the imaginary part of the meson polarization is
given in terms of the quantity
$\tilde{m}=m-2\,G_V\left(\rho_a-\rho_b \right)$, with $a,\,b$
standing for the flavors of quark and antiquark composing the
meson, and $m$ is its in medium mass. Since near the transition
point is $m \gg 2\,G_v\left(\rho_a-\rho_b\right)$ we consider in
the following only the case $\tilde{m}>0$. There are two sets of
conditions for the arising of an imaginary part coming from the
Dirac sea. The first one is composed by two simultaneous
inequalities
\begin{align}
0< \tilde{m}\leq M_b-M_a, \label{Imag1}
\end{align}
\begin{align}
\left[\tilde{m}^2-\left(M_a+M_b\right)^2\right]\,\left[\tilde{m}^2-\left(M_a-M_b\right)^2\right]<4\,\Lambda^2\,\tilde{m}^2.
\label{Imag2}
\end{align}
Eventually, one must take $\Lambda \rightarrow \infty$. For the
second set we have
\begin{align}
\tilde{m}>M_a+M_b, \label{Imag3}
\end{align}
\begin{align}
\left[\tilde{m}^2-\left(M_a+M_b\right)^2\right]\,\left[\tilde{m}^2-\left(M_a-M_b\right)^2\right]<4\,\Lambda^2\,\tilde{m}^2.
\label{Imag4}
\end{align}
An instability could also arise from the Fermi sea, in which case
we would have the same two sets of conditions but replacing
$\Lambda$ by $p_{Fa}$ in Eq.~(\ref{Imag2}) and by $p_{F b}$ in
Eq.~(\ref{Imag4}).

We have verified that the meson polarization insertions become
complex in the deconfined phase, indicating that the mesonic
excitations are unstable at extreme densities.

Now we focus on the results of the stellar matter case. For this
purpose, we use here the same parameters sets studied in
Ref.~\cite{MATTOS}, i.e., parametrizations of the PNJL0 model
given by (i) set~I: $G_V/G_s=0.15$, $a_3=-0.052$; (ii) set~II:
$G_V/G_s=0.25$, $a_3=-0.135$; (iii) set~III: $G_V/G_s=0.35$,
$a_3=-0.241$; and (iv) set~IV: $G_V/G_s=0.5$, $a_3=-0.458$. At the
low density regime we have used the hadronic DDH$\delta$ model and
both of them were connected through the Maxwell construction, in
this case giving rise to a hadron-quark phase transition with the
quark side composed by deconfined quarks, i.e., region in which
$\Phi>1$.

The behavior of the meson masses in SM is simpler, mainly because
the confined phase is described by a hadronic model.  The
interaction between nucleons and pseudoscalar mesons is motivated
by the chiral model \cite{KAPLAN}, and corresponds to a low order
approximation. Therefore it is adequate for a range of densities
below $ 2 \rho_0$, and extrapolations above such domain are highly
speculative. Moreover, our results are obtained in a mean field
approach excepting the pions which receive a special treatment due
to its nature of lowest Goldstone boson. The loss of the symmetry
in the light quark sector induces the splitting between the
members of a given iso-multiplet. However, beyond the
deconfinement transition they become almost degenerate and
unstable for all the considered cases.

\begin{figure}[!htb]
\centering
\includegraphics[scale=0.5]{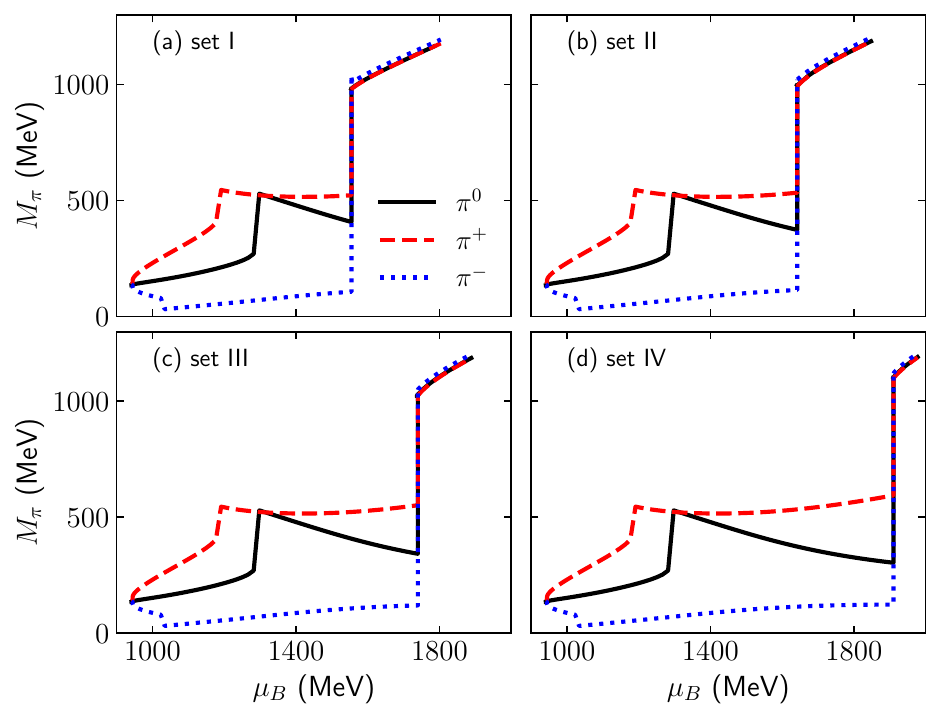}
\caption{Pion masses obtained from the PNJL0 model coupled to the
DDH$\delta$ one (stellar matter case) for different
parametrizations: (a) set I, (b) set II, (c) set III, and (d) set
IV.} \label{star-pions_mass}
\end{figure}

The pion masses are shown in Fig.~\ref{star-pions_mass}, for
several parameter sets.
As the increase of $G_V$ implies an increase of $|a_3|$, we find
that, in agreement with the result stated in the SFM case, the
confinement region extends to higher values of $\mu$ as $G_V$ is
increased. Furthermore, the mass gap at the transition point also
increases with $G_V$. In the hadronic matter the masses of the
multiplet are clearly separated for densities above $\rho_0/2$,
showing a smooth variation until a point of discontinuity. Beyond
such points the pions can exist only as unstable excitations.

\begin{figure}[!htb]
\centering
\includegraphics[scale=0.5]{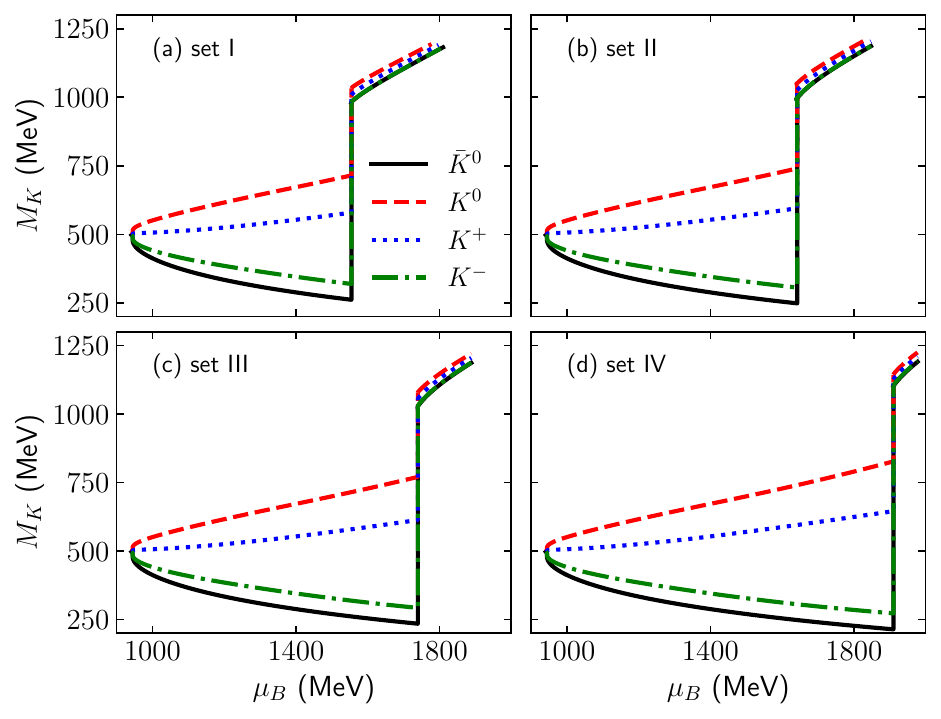}
\caption{Kaon masses obtained from the PNJL0 model coupled to the
DDH$\delta$ one (stellar matter case) for different
parametrizations: (a) set I, (b) set II, (c) set III, and (d) set
IV.} \label{star-kaons_mass}
\end{figure}

In regard of the kaons, there is a clear difference between the
$S=1$ and $S=-1$ components as can be seen in
Fig.~\ref{star-kaons_mass}.
In the first case, where $u, d$ flavors enter as antiparticles,  a
monotonous slight decrease with $\mu_B$ is obtained. In contrast,
the masses of both $K^+$ and $K^0$ show an increasing behavior.
These results are in qualitative agreement with those found by
\cite{MISHRA_SANYAL}.

\begin{figure}[!htb]
\centering
\includegraphics[scale=0.5]{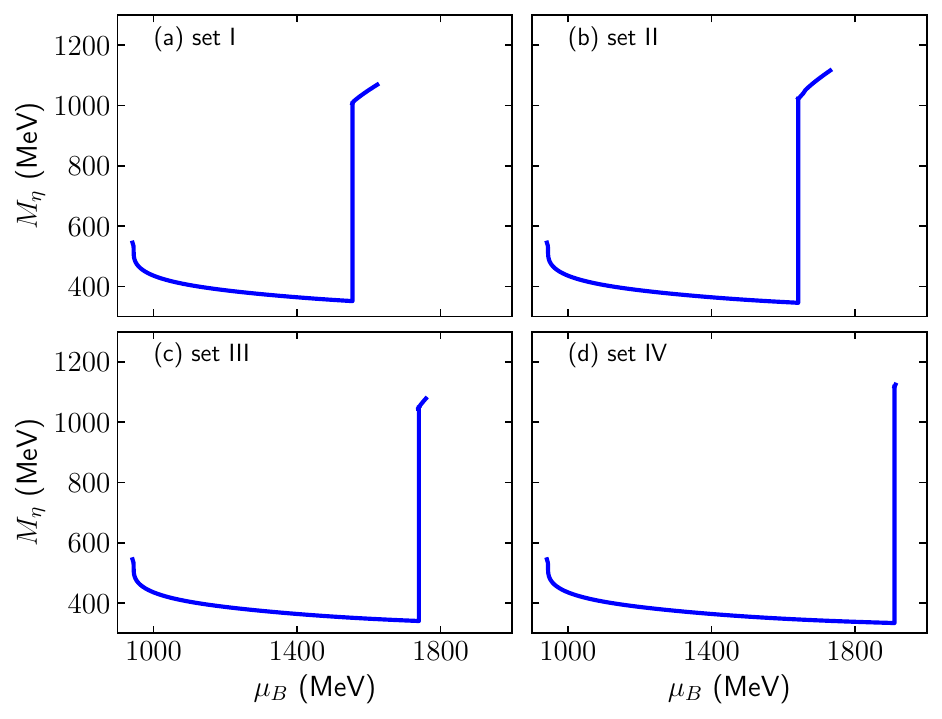}
\caption{$\eta$ mass obtained from the PNJL0 model coupled to the
DDH$\delta$ one (stellar matter case) for different
parametrizations: (a) set I, (b) set II, (c) set III, and (d) set
IV.} \label{star-etas_mass}
\end{figure}

The $\eta$ meson mass as function of the chemical potential is
displayed in Fig.~\ref{star-etas_mass}.
It experiences a strong decrease at very low densities in the
hadronic environment, before reaching a slight decreasing slope.
In the last regime the mass $M_\eta$ experiences a  strong
discontinuity at the threshold.

\begin{figure}[!htb]
\centering
\includegraphics[scale=0.5]{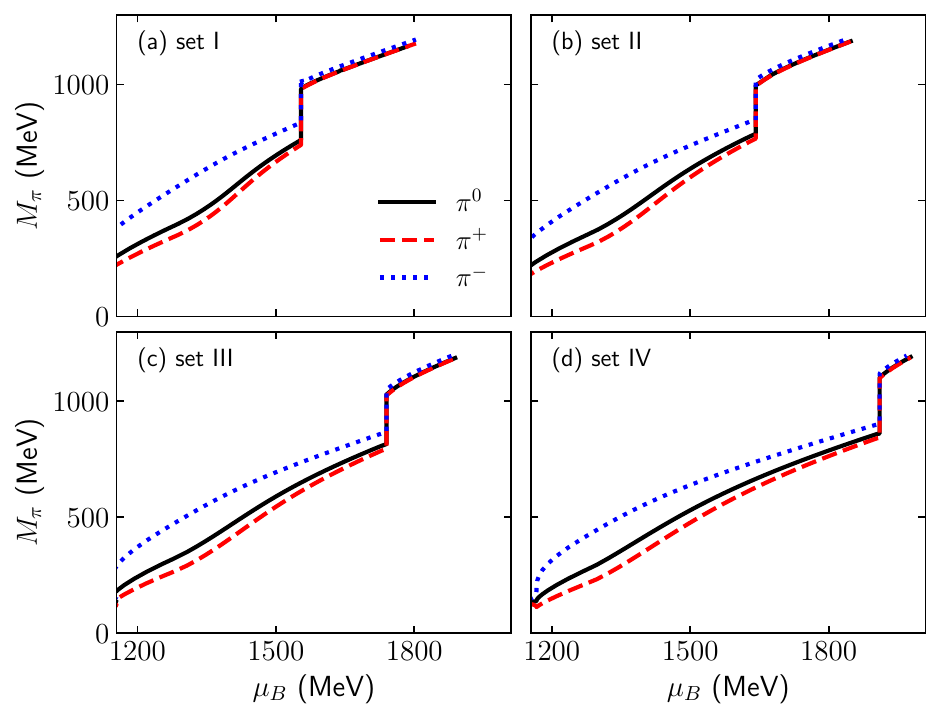}
\caption{Pion masses obtained from the PNJL0 model (stellar matter
case) for different parametrizations: (a)~set~I, (b)~set~II,
(c)~set~III, and (d)~set~IV.} \label{starpnjl0-pions_mass}
\end{figure}

Finally, for the sake of completeness, we present in
Figs.~\ref{starpnjl0-pions_mass}-\ref{starpnjl0-etas_mass} the
mesons masses calculated exclusively from the PNJL0 model in SM as
functions of the baryonic chemical potential~$\mu_B$. Although the
models used to describe densities below the deconfinement
threshold have dissimilar implementations, all of them are guided
by the ideas of chiral symmetry. This is the motivation to make a
daring contrast of the predictions for the in-medium meson masses
given by the PNJL0 and by the effective hadronic models, paying
special attention to the neighborhood of the transition point.

As the first comparison we consider the pion field in
Figs.~\ref{starpnjl0-pions_mass} and~\ref{star-pions_mass}.
Its effective mass has, within the domain $1050\, \text{MeV}
\leq\mu_B < 1200\, \text{MeV}$, a similar increasing trend in both
approaches. For $\mu_B$ slightly below $1200\, \text{MeV}$,
corresponding to a baryon density around $2.8\, \rho_0$, the
hadronic model shows a discontinuity in the $\pi^+$ mass, followed
by a plateau. The value $m_\pi \simeq 500\, \text{MeV}$ taken at
this point, marks the possible limit of applicability of the
model. In contrast, the PNJL0 predicts a monotonous increase, as a
consequence the mass gap at the transition point is comparatively
reduced. We expect the physical behavior for $\mu_B > 1200\,
\text{MeV}$ must lie between these qualitative descriptions. A
similar conclusion holds for the $\pi^0$ case.

\begin{figure}[!htb]
\centering
\includegraphics[scale=0.5]{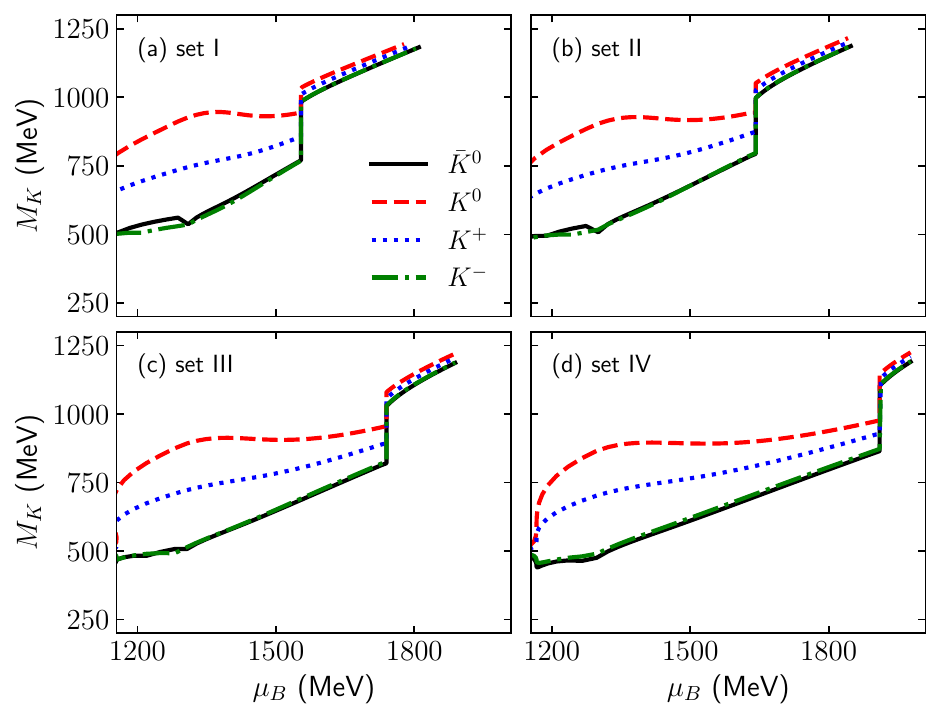}
\caption{Kaon masses obtained from the PNJL0 model (stellar matter
case) for different parametrizations: (a)~set~I, (b)~set~II,
(c)~set~III, and (d)~set~IV.} \label{starpnjl0-kaons_mass}
\end{figure}

As the next step we analyze the kaon masses in
Fig.~\ref{starpnjl0-kaons_mass}.
A contrast with the results shown in Fig.~\ref{star-kaons_mass}
evidences the smooth behavior, typical of the mean field approach,
obtained with the hadronic model, while the output of the PNJL0,
using a RPA, is more irregular. However, the monotonous increase
of the masses of the $S=1$ kaons is common for both calculations,
but more pronounced in the latter case. For the $S=-1$ kaons,
instead, these treatments give opposite behaviors. The main reason
is the negative contribution of the Weinberg-Tomozawa term of the
hadronic model~\cite{MISHRA_SANYAL} to the polarization of both
kaons. Moreover, the slope of the masses of the $S=-1$ kaons is
increased in the PNJL0 after the onset of the strange quarks to
the Fermi sea. In contrast, for the hadronic model there is no
strange degree of freedom since we are considering nuclear matter.
As for the pion masses, the faster increase with $\mu_B$ in the
low density regime of the PNJL0 leads to a lesser discontinuity of
the kaon masses at the deconfinement threshold.

To close this section we compare briefly the result for the $\eta$
mass in the PNJL0, as shown in Fig.~\ref{starpnjl0-etas_mass},
with the predictions of the hadronic model.
There is a clear opposition between the monotonous decrease found
in the latter and the definitely increasing output of the PNJL0
model. It must be bear in mind that the treatment of the quark
model includes the $\pi^0-\eta-\eta'$ mixing which is significant
as the $u-d$ symmetry is lost, as in fact occurs in the neutral
and beta equilibrated matter. In the hadronic model, instead, this
feature is disregarded and the parameter $\kappa$ is determined by
using the low-energy $\eta-N$ scattering data.

\section{Summary and concluding remarks}
\label{summary}

In this work we focus on the calculation of masses and couplings
related to the pseudoscalar $\pi$, $K$, and $\eta$ mesons. For
this purpose, we have used an effective three-flavor quark model
studied in Ref.~\cite{MATTOS} (PNJL0 model) in which the
phenomenology of the confinement/deconfinement phase transition at
zero temperature is taken into account by means of the traced
Polyakov loop, $\Phi$. In this model, the strength of the scalar,
vector, and 't Hoof channels are converted into $\Phi$-dependent
functions which vanish asymptotically in the deconfined region
when $\Phi\rightarrow 1$. This approach leads to a typical first
order phase transition, for which $\Phi$ is the order parameter,
even at zero temperature regime. This is a feature that is not
present in most of the PNJL models, that reduce to the NJL one at
$T=0$ limit. As performed in Ref.~\cite{MATTOS}, we consider the
PNJL0 model in two different scenarios, both at $T=0$. The first
one takes into account the symmetric flavor matter (SFM), for
which the quark chemical potentials are equal but with different
flavors distributed in different proportions (the quark densities
are different from each other). In the second case we focus on
quark matter under beta equilibrium conditions, i.e., with local
charge neutrality and chemical equilibrium. We named this scenario
as the stellar matter case (SM). For the latter case we describe
the hadron-quark phase transition by combining the PNJL0 model at
the quark sector, and the relativistic density dependent model
DDH$\delta$ at the hadronic side. The connection between both
sectors is made from the Maxwell construction, namely equality of
pressures and chemical potentials are imposed at the boundary of
both phases.

The mesons properties were calculated through the random phase
approximation, used in order to built the polarization insertion
$\Pi_a$ (related to the specific meson~$a$). This quantity is used
to extract the in-medium mesons masses as well as the effective
meson-quark-antiquark couplings, in this case identified from a
pole approximation. For the SFM scenario, $\Pi_a$ is exclusively
obtained from the PNJL0 model, from which Fermi momenta,
constituent masses and $\Phi$ are used as input to solve
Eqs~\eqref{zeromass} and~\eqref{coupling}. As a first result, we
verified that, for the pions, the deconfined phase (nonzero
solutions of $\Phi$) induces a significant increase in their
masses, and a drastic reduction in the pions-quark couplings
indicating an almost free pions regime. The gap in the masses
increase with the free parameters of the model, namely, the
strength of the vector channel $G_V$, and the constant $a_3$ that
regulates the magnitude of the Polyakov potential. For the sake of
comparison, we shown that our results coincide with the NJL ones
in the low chemical potential (or density) regime, region where
the confinement takes place ($\Phi=0$ solutions). The differences
arise exactly at the chemical potential where the $\Phi>0$
solutions appear. At this point, the almost linear increase
exhibited at the confined region presents a discontinuity.
Concerning the kaons, we verified that the $\bar{K}_0$ and $K^-$
coincide as well as those for $K_0$ and $K^+$ due to the symmetry
of the non-strange quarks. The non-monotonous dependence of their
masses in the NJL model is partially suppressed in the case of the
PNJL0 model. At the critical chemical potential, value at which it
is verified the beginning of the nonzero $\Phi$ solutions, it is
observed different behaviors for each kaon. For those with $S=1$,
the mass gap exhibited is around $100$~MeV while the $S=-1$ kaons
present a gap between $200$~MeV and $400$~MeV approximately.
Concerning the effective coupling for the kaons, we observe a
similar pattern in comparison with the pions case. It is possible
to clearly distinguish a phase of interacting kaons from the one
in which the system reaches a free phase. Similar features are
found for the $\eta$ and $\eta'$ results for masses and couplings.
Within our scheme we have found that if quark-antiquark bound
states persist beyond the deconfinement transition, they
correspond to heavy resonances. Moreover, this unstable mesons
interact weakly with its environment.

For the SM case we found similar results for the pion masses in
comparison to the SFM approach, namely, the critical baryonic
chemical potential increases as a function of the strength of the
vector channel $G_V$. For the kaons we verify that at the confined
region, governed by the hadronic model in SM, the mesons in which
$S=1$ ($K^-$ and $\bar{K}^0$) present a mass decreasing as a
function of $\mu_B$, in contrast with the $S=-1$ case, for which
both $M_{K^+}$ and $M_{K^0}$ increase with $\mu_B$. The decreasing
behavior of the mass is also observed for the $\eta$ meson at the
low baryonic chemical potential region. At the deconfined phase,
described by the PNJL0 model at beta equilibrium, a discontinuity
is observed, feature also registered for pions and kaons. The
comparison with the hadronic output, obtained by using a low order
expansion of the chiral interaction as a complement of the
DDH$\delta$ model, shows that the discontinuity of the masses of
the $\pi,\,K$ and $\eta$ mesons at the transition point is
considerable higher than in the PNJL0 model. Furthermore, we have
found a qualitative agreement for the masses of the $K^0$ and
$K^+$ fields. For the pions this similitude extends over baryon
densities $\rho/\rho_0< 2.8$, while for the remaining $\bar{K}^0,
K^-$, and $\eta$ mesons, the results are not compatible.

\section*{Acknowledgements}
R.~M.~A. acknowledges the support given by the CONICET of
Argentina under the grant PIP-11220200102081CO. O.~L. is member of
the project INCT-FNA Proc. No. 464898/2014-5 and acknowledges the
support provided by both, Conselho Nacional de Desenvolvimento
Cient\'ifico e Tecnol\'ogico (CNPq) under Grant No. 312410/2020-4,
and Funda\c{c}\~ao de Amparo \`a Pesquisa do Estado de S\~ao Paulo
(FAPESP) under and Grant No. 2022/03575-3 (BPE).

\bibliographystyle{apsrev4-2}
\bibliography{references}

\begin{thebibliography}{59}%
\makeatletter
\providecommand \@ifxundefined [1]{%
 \@ifx{#1\undefined}
}%
\providecommand \@ifnum [1]{%
 \ifnum #1\expandafter \@firstoftwo
 \else \expandafter \@secondoftwo
 \fi
}%
\providecommand \@ifx [1]{%
 \ifx #1\expandafter \@firstoftwo
 \else \expandafter \@secondoftwo
 \fi
}%
\providecommand \natexlab [1]{#1}%
\providecommand \enquote  [1]{``#1''}%
\providecommand \bibnamefont  [1]{#1}%
\providecommand \bibfnamefont [1]{#1}%
\providecommand \citenamefont [1]{#1}%
\providecommand \href@noop [0]{\@secondoftwo}%
\providecommand \href [0]{\begingroup \@sanitize@url \@href}%
\providecommand \@href[1]{\@@startlink{#1}\@@href}%
\providecommand \@@href[1]{\endgroup#1\@@endlink}%
\providecommand \@sanitize@url [0]{\catcode `\\12\catcode `\$12\catcode
  `\&12\catcode `\#12\catcode `\^12\catcode `\_12\catcode `\%12\relax}%
\providecommand \@@startlink[1]{}%
\providecommand \@@endlink[0]{}%
\providecommand \url  [0]{\begingroup\@sanitize@url \@url }%
\providecommand \@url [1]{\endgroup\@href {#1}{\urlprefix }}%
\providecommand \urlprefix  [0]{URL }%
\providecommand \Eprint [0]{\href }%
\providecommand \doibase [0]{https://doi.org/}%
\providecommand \selectlanguage [0]{\@gobble}%
\providecommand \bibinfo  [0]{\@secondoftwo}%
\providecommand \bibfield  [0]{\@secondoftwo}%
\providecommand \translation [1]{[#1]}%
\providecommand \BibitemOpen [0]{}%
\providecommand \bibitemStop [0]{}%
\providecommand \bibitemNoStop [0]{.\EOS\space}%
\providecommand \EOS [0]{\spacefactor3000\relax}%
\providecommand \BibitemShut  [1]{\csname bibitem#1\endcsname}%
\let\auto@bib@innerbib\@empty
\bibitem [{\citenamefont {Hayano}\ and\ \citenamefont
  {Hatsuda}(2010)}]{hayano}%
  \BibitemOpen
  \bibfield  {author} {\bibinfo {author} {\bibfnamefont {R.~S.}\ \bibnamefont
  {Hayano}}\ and\ \bibinfo {author} {\bibfnamefont {T.}~\bibnamefont
  {Hatsuda}},\ }\href {https://doi.org/10.1103/RevModPhys.82.2949} {\bibfield
  {journal} {\bibinfo  {journal} {Rev. Mod. Phys.}\ }\textbf {\bibinfo {volume}
  {82}},\ \bibinfo {pages} {2949} (\bibinfo {year} {2010})}\BibitemShut
  {NoStop}%
\bibitem [{\citenamefont {Hansen}\ \emph {et~al.}(2007)\citenamefont {Hansen},
  \citenamefont {Alberico}, \citenamefont {Beraudo}, \citenamefont {Molinari},
  \citenamefont {Nardi},\ and\ \citenamefont {Ratti}}]{HANSEN}%
  \BibitemOpen
  \bibfield  {author} {\bibinfo {author} {\bibfnamefont {H.}~\bibnamefont
  {Hansen}}, \bibinfo {author} {\bibfnamefont {W.~M.}\ \bibnamefont
  {Alberico}}, \bibinfo {author} {\bibfnamefont {A.}~\bibnamefont {Beraudo}},
  \bibinfo {author} {\bibfnamefont {A.}~\bibnamefont {Molinari}}, \bibinfo
  {author} {\bibfnamefont {M.}~\bibnamefont {Nardi}},\ and\ \bibinfo {author}
  {\bibfnamefont {C.}~\bibnamefont {Ratti}},\ }\href
  {https://doi.org/10.1103/PhysRevD.75.065004} {\bibfield  {journal} {\bibinfo
  {journal} {Phys. Rev. D}\ }\textbf {\bibinfo {volume} {75}},\ \bibinfo
  {pages} {065004} (\bibinfo {year} {2007})}\BibitemShut {NoStop}%
\bibitem [{\citenamefont {Fu}\ and\ \citenamefont {Liu}(2009)}]{FU_LIU}%
  \BibitemOpen
  \bibfield  {author} {\bibinfo {author} {\bibfnamefont {W.-j.}\ \bibnamefont
  {Fu}}\ and\ \bibinfo {author} {\bibfnamefont {Y.-x.}\ \bibnamefont {Liu}},\
  }\href {https://doi.org/10.1103/PhysRevD.79.074011} {\bibfield  {journal}
  {\bibinfo  {journal} {Phys. Rev. D}\ }\textbf {\bibinfo {volume} {79}},\
  \bibinfo {pages} {074011} (\bibinfo {year} {2009})}\BibitemShut {NoStop}%
\bibitem [{\citenamefont {Abuki}\ \emph {et~al.}(2008)\citenamefont {Abuki},
  \citenamefont {Ciminale}, \citenamefont {Gatto}, \citenamefont {Ippolito},
  \citenamefont {Nardulli},\ and\ \citenamefont {Ruggieri}}]{ABUKI}%
  \BibitemOpen
  \bibfield  {author} {\bibinfo {author} {\bibfnamefont {H.}~\bibnamefont
  {Abuki}}, \bibinfo {author} {\bibfnamefont {M.}~\bibnamefont {Ciminale}},
  \bibinfo {author} {\bibfnamefont {R.}~\bibnamefont {Gatto}}, \bibinfo
  {author} {\bibfnamefont {N.~D.}\ \bibnamefont {Ippolito}}, \bibinfo {author}
  {\bibfnamefont {G.}~\bibnamefont {Nardulli}},\ and\ \bibinfo {author}
  {\bibfnamefont {M.}~\bibnamefont {Ruggieri}},\ }\href
  {https://doi.org/10.1103/PhysRevD.78.014002} {\bibfield  {journal} {\bibinfo
  {journal} {Phys. Rev. D}\ }\textbf {\bibinfo {volume} {78}},\ \bibinfo
  {pages} {014002} (\bibinfo {year} {2008})}\BibitemShut {NoStop}%
\bibitem [{\citenamefont {Yamazaki}\ and\ \citenamefont
  {Matsui}(2014)}]{YAMAZAKI}%
  \BibitemOpen
  \bibfield  {author} {\bibinfo {author} {\bibfnamefont {K.}~\bibnamefont
  {Yamazaki}}\ and\ \bibinfo {author} {\bibfnamefont {T.}~\bibnamefont
  {Matsui}},\ }\href
  {https://doi.org/https://doi.org/10.1016/j.nuclphysa.2013.12.010} {\bibfield
  {journal} {\bibinfo  {journal} {Nuclear Physics A}\ }\textbf {\bibinfo
  {volume} {922}},\ \bibinfo {pages} {237} (\bibinfo {year}
  {2014})}\BibitemShut {NoStop}%
\bibitem [{\citenamefont {Costa}\ \emph {et~al.}(2009)\citenamefont {Costa},
  \citenamefont {Ruivo}, \citenamefont {de~Sousa}, \citenamefont {Hansen},\
  and\ \citenamefont {Alberico}}]{COSTA2}%
  \BibitemOpen
  \bibfield  {author} {\bibinfo {author} {\bibfnamefont {P.}~\bibnamefont
  {Costa}}, \bibinfo {author} {\bibfnamefont {M.~C.}\ \bibnamefont {Ruivo}},
  \bibinfo {author} {\bibfnamefont {C.~A.}\ \bibnamefont {de~Sousa}}, \bibinfo
  {author} {\bibfnamefont {H.}~\bibnamefont {Hansen}},\ and\ \bibinfo {author}
  {\bibfnamefont {W.~M.}\ \bibnamefont {Alberico}},\ }\href
  {https://doi.org/10.1103/PhysRevD.79.116003} {\bibfield  {journal} {\bibinfo
  {journal} {Phys. Rev. D}\ }\textbf {\bibinfo {volume} {79}},\ \bibinfo
  {pages} {116003} (\bibinfo {year} {2009})}\BibitemShut {NoStop}%
\bibitem [{\citenamefont {Gupta}\ and\ \citenamefont {Tiwari}(2010)}]{GUPTA}%
  \BibitemOpen
  \bibfield  {author} {\bibinfo {author} {\bibfnamefont {U.~S.}\ \bibnamefont
  {Gupta}}\ and\ \bibinfo {author} {\bibfnamefont {V.~K.}\ \bibnamefont
  {Tiwari}},\ }\href {https://doi.org/10.1103/PhysRevD.81.054019} {\bibfield
  {journal} {\bibinfo  {journal} {Phys. Rev. D}\ }\textbf {\bibinfo {volume}
  {81}},\ \bibinfo {pages} {054019} (\bibinfo {year} {2010})}\BibitemShut
  {NoStop}%
\bibitem [{\citenamefont {Ratti}\ \emph {et~al.}(2012)\citenamefont {Ratti},
  \citenamefont {Bellwied}, \citenamefont {Cristoforetti},\ and\ \citenamefont
  {Barbaro}}]{RATTI}%
  \BibitemOpen
  \bibfield  {author} {\bibinfo {author} {\bibfnamefont {C.}~\bibnamefont
  {Ratti}}, \bibinfo {author} {\bibfnamefont {R.}~\bibnamefont {Bellwied}},
  \bibinfo {author} {\bibfnamefont {M.}~\bibnamefont {Cristoforetti}},\ and\
  \bibinfo {author} {\bibfnamefont {M.}~\bibnamefont {Barbaro}},\ }\href
  {https://doi.org/10.1103/PhysRevD.85.014004} {\bibfield  {journal} {\bibinfo
  {journal} {Phys. Rev. D}\ }\textbf {\bibinfo {volume} {85}},\ \bibinfo
  {pages} {014004} (\bibinfo {year} {2012})}\BibitemShut {NoStop}%
\bibitem [{\citenamefont {Bellwied}\ \emph {et~al.}(2013)\citenamefont
  {Bellwied}, \citenamefont {Borsanyi}, \citenamefont {Fodor}, \citenamefont
  {Katz},\ and\ \citenamefont {Ratti}}]{BELLWIED}%
  \BibitemOpen
  \bibfield  {author} {\bibinfo {author} {\bibfnamefont {R.}~\bibnamefont
  {Bellwied}}, \bibinfo {author} {\bibfnamefont {S.}~\bibnamefont {Borsanyi}},
  \bibinfo {author} {\bibfnamefont {Z.}~\bibnamefont {Fodor}}, \bibinfo
  {author} {\bibfnamefont {S.~D.}\ \bibnamefont {Katz}},\ and\ \bibinfo
  {author} {\bibfnamefont {C.}~\bibnamefont {Ratti}},\ }\href
  {https://doi.org/10.1103/PhysRevLett.111.202302} {\bibfield  {journal}
  {\bibinfo  {journal} {Phys. Rev. Lett.}\ }\textbf {\bibinfo {volume} {111}},\
  \bibinfo {pages} {202302} (\bibinfo {year} {2013})}\BibitemShut {NoStop}%
\bibitem [{\citenamefont {{Bugaev}}\ \emph {et~al.}(2016)\citenamefont
  {{Bugaev}}, \citenamefont {{Sagun}}, \citenamefont {{Ivanytskyi}},
  \citenamefont {{Oliinychenko}}, \citenamefont {{Ilgenfritz}}, \citenamefont
  {{Nikonov}}, \citenamefont {{Taranenko}},\ and\ \citenamefont
  {{Zinovjev}}}]{BUGAEV}%
  \BibitemOpen
  \bibfield  {author} {\bibinfo {author} {\bibfnamefont {K.~A.}\ \bibnamefont
  {{Bugaev}}}, \bibinfo {author} {\bibfnamefont {V.~V.}\ \bibnamefont
  {{Sagun}}}, \bibinfo {author} {\bibfnamefont {A.~I.}\ \bibnamefont
  {{Ivanytskyi}}}, \bibinfo {author} {\bibfnamefont {D.~R.}\ \bibnamefont
  {{Oliinychenko}}}, \bibinfo {author} {\bibfnamefont {E.~M.}\ \bibnamefont
  {{Ilgenfritz}}}, \bibinfo {author} {\bibfnamefont {E.~G.}\ \bibnamefont
  {{Nikonov}}}, \bibinfo {author} {\bibfnamefont {A.~V.}\ \bibnamefont
  {{Taranenko}}},\ and\ \bibinfo {author} {\bibfnamefont {G.~M.}\ \bibnamefont
  {{Zinovjev}}},\ }\href {https://doi.org/10.1140/epja/i2016-16227-6}
  {\bibfield  {journal} {\bibinfo  {journal} {European Physical Journal A}\
  }\textbf {\bibinfo {volume} {52}},\ \bibinfo {eid} {227} (\bibinfo {year}
  {2016})}\BibitemShut {NoStop}%
\bibitem [{\citenamefont {{Di Toro}}\ \emph {et~al.}(2016)\citenamefont {{Di
  Toro}}, \citenamefont {{Colonna}}, \citenamefont {{Greco}},\ and\
  \citenamefont {{Shao}}}]{DITORO}%
  \BibitemOpen
  \bibfield  {author} {\bibinfo {author} {\bibfnamefont {M.}~\bibnamefont {{Di
  Toro}}}, \bibinfo {author} {\bibfnamefont {M.}~\bibnamefont {{Colonna}}},
  \bibinfo {author} {\bibfnamefont {V.}~\bibnamefont {{Greco}}},\ and\ \bibinfo
  {author} {\bibfnamefont {G.-Y.}\ \bibnamefont {{Shao}}},\ }\href
  {https://doi.org/10.1140/epja/i2016-16224-9} {\bibfield  {journal} {\bibinfo
  {journal} {European Physical Journal A}\ }\textbf {\bibinfo {volume} {52}},\
  \bibinfo {eid} {224} (\bibinfo {year} {2016})}\BibitemShut {NoStop}%
\bibitem [{\citenamefont {Glendenning}(1992)}]{GLENDENNING}%
  \BibitemOpen
  \bibfield  {author} {\bibinfo {author} {\bibfnamefont {N.~K.}\ \bibnamefont
  {Glendenning}},\ }\href {https://doi.org/10.1103/PhysRevD.46.1274} {\bibfield
   {journal} {\bibinfo  {journal} {Phys. Rev. D}\ }\textbf {\bibinfo {volume}
  {46}},\ \bibinfo {pages} {1274} (\bibinfo {year} {1992})}\BibitemShut
  {NoStop}%
\bibitem [{\citenamefont {{Annala}}\ \emph {et~al.}(2020)\citenamefont
  {{Annala}}, \citenamefont {{Gorda}}, \citenamefont {{Kurkela}}, \citenamefont
  {{N{\"a}ttil{\"a}}},\ and\ \citenamefont {{Vuorinen}}}]{nat1}%
  \BibitemOpen
  \bibfield  {author} {\bibinfo {author} {\bibfnamefont {E.}~\bibnamefont
  {{Annala}}}, \bibinfo {author} {\bibfnamefont {T.}~\bibnamefont {{Gorda}}},
  \bibinfo {author} {\bibfnamefont {A.}~\bibnamefont {{Kurkela}}}, \bibinfo
  {author} {\bibfnamefont {J.}~\bibnamefont {{N{\"a}ttil{\"a}}}},\ and\
  \bibinfo {author} {\bibfnamefont {A.}~\bibnamefont {{Vuorinen}}},\ }\href
  {https://doi.org/10.1038/s41567-020-0914-9} {\bibfield  {journal} {\bibinfo
  {journal} {Nature Physics}\ }\textbf {\bibinfo {volume} {16}},\ \bibinfo
  {pages} {907} (\bibinfo {year} {2020})}\BibitemShut {NoStop}%
\bibitem [{\citenamefont {{Annala}}\ \emph {et~al.}(2023)\citenamefont
  {{Annala}}, \citenamefont {{Gorda}}, \citenamefont {{Hirvonen}},
  \citenamefont {{Komoltsev}}, \citenamefont {{Kurkela}}, \citenamefont
  {{N{\"a}ttil{\"a}}},\ and\ \citenamefont {{Vuorinen}}}]{nat2}%
  \BibitemOpen
  \bibfield  {author} {\bibinfo {author} {\bibfnamefont {E.}~\bibnamefont
  {{Annala}}}, \bibinfo {author} {\bibfnamefont {T.}~\bibnamefont {{Gorda}}},
  \bibinfo {author} {\bibfnamefont {J.}~\bibnamefont {{Hirvonen}}}, \bibinfo
  {author} {\bibfnamefont {O.}~\bibnamefont {{Komoltsev}}}, \bibinfo {author}
  {\bibfnamefont {A.}~\bibnamefont {{Kurkela}}}, \bibinfo {author}
  {\bibfnamefont {J.}~\bibnamefont {{N{\"a}ttil{\"a}}}},\ and\ \bibinfo
  {author} {\bibfnamefont {A.}~\bibnamefont {{Vuorinen}}},\ }\href
  {https://doi.org/10.1038/s41467-023-44051-y} {\bibfield  {journal} {\bibinfo
  {journal} {Nature Communications}\ }\textbf {\bibinfo {volume} {14}},\
  \bibinfo {eid} {8451} (\bibinfo {year} {2023})}\BibitemShut {NoStop}%
\bibitem [{\citenamefont {Barducci}\ \emph {et~al.}(2005)\citenamefont
  {Barducci}, \citenamefont {Casalbuoni}, \citenamefont {Pettini},\ and\
  \citenamefont {Ravagli}}]{BARDUCCI}%
  \BibitemOpen
  \bibfield  {author} {\bibinfo {author} {\bibfnamefont {A.}~\bibnamefont
  {Barducci}}, \bibinfo {author} {\bibfnamefont {R.}~\bibnamefont
  {Casalbuoni}}, \bibinfo {author} {\bibfnamefont {G.}~\bibnamefont
  {Pettini}},\ and\ \bibinfo {author} {\bibfnamefont {L.}~\bibnamefont
  {Ravagli}},\ }\href {https://doi.org/10.1103/PhysRevD.71.016011} {\bibfield
  {journal} {\bibinfo  {journal} {Phys. Rev. D}\ }\textbf {\bibinfo {volume}
  {71}},\ \bibinfo {pages} {016011} (\bibinfo {year} {2005})}\BibitemShut
  {NoStop}%
\bibitem [{\citenamefont {Sch\"afer}(2000)}]{SCHAFER}%
  \BibitemOpen
  \bibfield  {author} {\bibinfo {author} {\bibfnamefont {T.}~\bibnamefont
  {Sch\"afer}},\ }\href {https://doi.org/10.1103/PhysRevLett.85.5531}
  {\bibfield  {journal} {\bibinfo  {journal} {Phys. Rev. Lett.}\ }\textbf
  {\bibinfo {volume} {85}},\ \bibinfo {pages} {5531} (\bibinfo {year}
  {2000})}\BibitemShut {NoStop}%
\bibitem [{\citenamefont {Basler}\ and\ \citenamefont
  {Buballa}(2010)}]{BASLER}%
  \BibitemOpen
  \bibfield  {author} {\bibinfo {author} {\bibfnamefont {H.}~\bibnamefont
  {Basler}}\ and\ \bibinfo {author} {\bibfnamefont {M.}~\bibnamefont
  {Buballa}},\ }\href {https://doi.org/10.1103/PhysRevD.81.054033} {\bibfield
  {journal} {\bibinfo  {journal} {Phys. Rev. D}\ }\textbf {\bibinfo {volume}
  {81}},\ \bibinfo {pages} {054033} (\bibinfo {year} {2010})}\BibitemShut
  {NoStop}%
\bibitem [{\citenamefont {Steinheimer}\ \emph {et~al.}(2011)\citenamefont
  {Steinheimer}, \citenamefont {Schramm},\ and\ \citenamefont
  {Stöcker}}]{STEINHEIMER}%
  \BibitemOpen
  \bibfield  {author} {\bibinfo {author} {\bibfnamefont {J.}~\bibnamefont
  {Steinheimer}}, \bibinfo {author} {\bibfnamefont {S.}~\bibnamefont
  {Schramm}},\ and\ \bibinfo {author} {\bibfnamefont {H.}~\bibnamefont
  {Stöcker}},\ }\href {https://doi.org/10.1088/0954-3899/38/3/035001}
  {\bibfield  {journal} {\bibinfo  {journal} {Journal of Physics G: Nuclear and
  Particle Physics}\ }\textbf {\bibinfo {volume} {38}},\ \bibinfo {pages}
  {035001} (\bibinfo {year} {2011})}\BibitemShut {NoStop}%
\bibitem [{\citenamefont {Beni\ifmmode~\acute{c}\else \'{c}\fi{}}\ \emph
  {et~al.}(2015)\citenamefont {Beni\ifmmode~\acute{c}\else \'{c}\fi{}},
  \citenamefont {Mishustin},\ and\ \citenamefont {Sasaki}}]{BENIC}%
  \BibitemOpen
  \bibfield  {author} {\bibinfo {author} {\bibfnamefont {S.}~\bibnamefont
  {Beni\ifmmode~\acute{c}\else \'{c}\fi{}}}, \bibinfo {author} {\bibfnamefont
  {I.}~\bibnamefont {Mishustin}},\ and\ \bibinfo {author} {\bibfnamefont
  {C.}~\bibnamefont {Sasaki}},\ }\href
  {https://doi.org/10.1103/PhysRevD.91.125034} {\bibfield  {journal} {\bibinfo
  {journal} {Phys. Rev. D}\ }\textbf {\bibinfo {volume} {91}},\ \bibinfo
  {pages} {125034} (\bibinfo {year} {2015})}\BibitemShut {NoStop}%
\bibitem [{\citenamefont {Mattos}\ \emph {et~al.}(2021)\citenamefont {Mattos},
  \citenamefont {Frederico}, \citenamefont {Lenzi}, \citenamefont {Dutra},\
  and\ \citenamefont {Louren\ifmmode~\mbox{\c{c}}\else \c{c}\fi{}o}}]{MATTOS}%
  \BibitemOpen
  \bibfield  {author} {\bibinfo {author} {\bibfnamefont {O.~A.}\ \bibnamefont
  {Mattos}}, \bibinfo {author} {\bibfnamefont {T.}~\bibnamefont {Frederico}},
  \bibinfo {author} {\bibfnamefont {C.~H.}\ \bibnamefont {Lenzi}}, \bibinfo
  {author} {\bibfnamefont {M.}~\bibnamefont {Dutra}},\ and\ \bibinfo {author}
  {\bibfnamefont {O.}~\bibnamefont {Louren\ifmmode~\mbox{\c{c}}\else
  \c{c}\fi{}o}},\ }\href {https://doi.org/10.1103/PhysRevD.104.116001}
  {\bibfield  {journal} {\bibinfo  {journal} {Phys. Rev. D}\ }\textbf {\bibinfo
  {volume} {104}},\ \bibinfo {pages} {116001} (\bibinfo {year}
  {2021})}\BibitemShut {NoStop}%
\bibitem [{\citenamefont {Mattos}\ \emph {et~al.}(2019)\citenamefont {Mattos},
  \citenamefont {Lourenço},\ and\ \citenamefont {Frederico}}]{Mattos_2019}%
  \BibitemOpen
  \bibfield  {author} {\bibinfo {author} {\bibfnamefont {O.~A.}\ \bibnamefont
  {Mattos}}, \bibinfo {author} {\bibfnamefont {O.}~\bibnamefont {Lourenço}},\
  and\ \bibinfo {author} {\bibfnamefont {T.}~\bibnamefont {Frederico}},\ }\href
  {https://doi.org/10.1088/1742-6596/1291/1/012031} {\bibfield  {journal}
  {\bibinfo  {journal} {Journal of Physics: Conference Series}\ }\textbf
  {\bibinfo {volume} {1291}},\ \bibinfo {pages} {012031} (\bibinfo {year}
  {2019})}\BibitemShut {NoStop}%
\bibitem [{\citenamefont {{Mattos}}\ \emph {et~al.}(2021)\citenamefont
  {{Mattos}}, \citenamefont {{Frederico}},\ and\ \citenamefont
  {{Louren{\c{c}}o}}}]{Mattos_2021}%
  \BibitemOpen
  \bibfield  {author} {\bibinfo {author} {\bibfnamefont {O.~A.}\ \bibnamefont
  {{Mattos}}}, \bibinfo {author} {\bibfnamefont {T.}~\bibnamefont
  {{Frederico}}},\ and\ \bibinfo {author} {\bibfnamefont {O.}~\bibnamefont
  {{Louren{\c{c}}o}}},\ }\href
  {https://doi.org/10.1140/epjc/s10052-021-08827-0} {\bibfield  {journal}
  {\bibinfo  {journal} {European Physical Journal C}\ }\textbf {\bibinfo
  {volume} {81}},\ \bibinfo {eid} {24} (\bibinfo {year} {2021})}\BibitemShut
  {NoStop}%
\bibitem [{\citenamefont {Pons}\ \emph {et~al.}(2001)\citenamefont {Pons},
  \citenamefont {Miralles}, \citenamefont {Prakash},\ and\ \citenamefont
  {Lattimer}}]{PONS}%
  \BibitemOpen
  \bibfield  {author} {\bibinfo {author} {\bibfnamefont {J.~A.}\ \bibnamefont
  {Pons}}, \bibinfo {author} {\bibfnamefont {J.~A.}\ \bibnamefont {Miralles}},
  \bibinfo {author} {\bibfnamefont {M.}~\bibnamefont {Prakash}},\ and\ \bibinfo
  {author} {\bibfnamefont {J.~M.}\ \bibnamefont {Lattimer}},\ }\href
  {https://doi.org/10.1086/320642} {\bibfield  {journal} {\bibinfo  {journal}
  {The Astrophysical Journal}\ }\textbf {\bibinfo {volume} {553}},\ \bibinfo
  {pages} {382} (\bibinfo {year} {2001})}\BibitemShut {NoStop}%
\bibitem [{\citenamefont {Vijayan}\ \emph {et~al.}(2023)\citenamefont
  {Vijayan}, \citenamefont {Rahman}, \citenamefont {Bauswein}, \citenamefont
  {Mart\'{\i}nez-Pinedo},\ and\ \citenamefont {Arbina}}]{VIJAYAN}%
  \BibitemOpen
  \bibfield  {author} {\bibinfo {author} {\bibfnamefont {V.}~\bibnamefont
  {Vijayan}}, \bibinfo {author} {\bibfnamefont {N.}~\bibnamefont {Rahman}},
  \bibinfo {author} {\bibfnamefont {A.}~\bibnamefont {Bauswein}}, \bibinfo
  {author} {\bibfnamefont {G.}~\bibnamefont {Mart\'{\i}nez-Pinedo}},\ and\
  \bibinfo {author} {\bibfnamefont {I.~L.}\ \bibnamefont {Arbina}},\ }\href
  {https://doi.org/10.1103/PhysRevD.108.023020} {\bibfield  {journal} {\bibinfo
   {journal} {Phys. Rev. D}\ }\textbf {\bibinfo {volume} {108}},\ \bibinfo
  {pages} {023020} (\bibinfo {year} {2023})}\BibitemShut {NoStop}%
\bibitem [{\citenamefont {Yakovlev}\ and\ \citenamefont
  {Pethick}(2004)}]{YAKOVLEV}%
  \BibitemOpen
  \bibfield  {author} {\bibinfo {author} {\bibfnamefont {D.}~\bibnamefont
  {Yakovlev}}\ and\ \bibinfo {author} {\bibfnamefont {C.}~\bibnamefont
  {Pethick}},\ }\href {https://doi.org/10.1146/annurev.astro.42.053102.134013}
  {\bibfield  {journal} {\bibinfo  {journal} {Annual Review of Astronomy and
  Astrophysics}\ }\textbf {\bibinfo {volume} {42}},\ \bibinfo {pages} {169}
  (\bibinfo {year} {2004})}\BibitemShut {NoStop}%
\bibitem [{\citenamefont {Friedman}\ and\ \citenamefont
  {Gal}(2007)}]{FRIEDMAN}%
  \BibitemOpen
  \bibfield  {author} {\bibinfo {author} {\bibfnamefont {E.}~\bibnamefont
  {Friedman}}\ and\ \bibinfo {author} {\bibfnamefont {A.}~\bibnamefont {Gal}},\
  }\href {https://doi.org/https://doi.org/10.1016/j.physrep.2007.08.002}
  {\bibfield  {journal} {\bibinfo  {journal} {Physics Reports}\ }\textbf
  {\bibinfo {volume} {452}},\ \bibinfo {pages} {89} (\bibinfo {year}
  {2007})}\BibitemShut {NoStop}%
\bibitem [{\citenamefont {Muto}\ \emph {et~al.}(2009)\citenamefont {Muto},
  \citenamefont {Maruyama},\ and\ \citenamefont {Tatsumi}}]{MUTO}%
  \BibitemOpen
  \bibfield  {author} {\bibinfo {author} {\bibfnamefont {T.}~\bibnamefont
  {Muto}}, \bibinfo {author} {\bibfnamefont {T.}~\bibnamefont {Maruyama}},\
  and\ \bibinfo {author} {\bibfnamefont {T.}~\bibnamefont {Tatsumi}},\ }\href
  {https://doi.org/10.1103/PhysRevC.79.035207} {\bibfield  {journal} {\bibinfo
  {journal} {Phys. Rev. C}\ }\textbf {\bibinfo {volume} {79}},\ \bibinfo
  {pages} {035207} (\bibinfo {year} {2009})}\BibitemShut {NoStop}%
\bibitem [{\citenamefont {Cieplý}\ \emph {et~al.}(2014)\citenamefont
  {Cieplý}, \citenamefont {Friedman}, \citenamefont {Gal},\ and\ \citenamefont
  {Mareš}}]{CIEPLY}%
  \BibitemOpen
  \bibfield  {author} {\bibinfo {author} {\bibfnamefont {A.}~\bibnamefont
  {Cieplý}}, \bibinfo {author} {\bibfnamefont {E.}~\bibnamefont {Friedman}},
  \bibinfo {author} {\bibfnamefont {A.}~\bibnamefont {Gal}},\ and\ \bibinfo
  {author} {\bibfnamefont {J.}~\bibnamefont {Mareš}},\ }\href
  {https://doi.org/https://doi.org/10.1016/j.nuclphysa.2014.02.007} {\bibfield
  {journal} {\bibinfo  {journal} {Nuclear Physics A}\ }\textbf {\bibinfo
  {volume} {925}},\ \bibinfo {pages} {126} (\bibinfo {year}
  {2014})}\BibitemShut {NoStop}%
\bibitem [{\citenamefont {Song}\ \emph {et~al.}(2008)\citenamefont {Song},
  \citenamefont {Zhong}, \citenamefont {Li},\ and\ \citenamefont
  {Ning}}]{SONG}%
  \BibitemOpen
  \bibfield  {author} {\bibinfo {author} {\bibfnamefont {C.~Y.}\ \bibnamefont
  {Song}}, \bibinfo {author} {\bibfnamefont {X.~H.}\ \bibnamefont {Zhong}},
  \bibinfo {author} {\bibfnamefont {L.}~\bibnamefont {Li}},\ and\ \bibinfo
  {author} {\bibfnamefont {P.~Z.}\ \bibnamefont {Ning}},\ }\href
  {https://doi.org/10.1209/0295-5075/81/42002} {\bibfield  {journal} {\bibinfo
  {journal} {Europhysics Letters}\ }\textbf {\bibinfo {volume} {81}},\ \bibinfo
  {pages} {42002} (\bibinfo {year} {2008})}\BibitemShut {NoStop}%
\bibitem [{\citenamefont {Nagahiro}\ \emph {et~al.}(2009)\citenamefont
  {Nagahiro}, \citenamefont {Jido},\ and\ \citenamefont
  {Hirenzaki}}]{NAGAHIRO}%
  \BibitemOpen
  \bibfield  {author} {\bibinfo {author} {\bibfnamefont {H.}~\bibnamefont
  {Nagahiro}}, \bibinfo {author} {\bibfnamefont {D.}~\bibnamefont {Jido}},\
  and\ \bibinfo {author} {\bibfnamefont {S.}~\bibnamefont {Hirenzaki}},\ }\href
  {https://doi.org/10.1103/PhysRevC.80.025205} {\bibfield  {journal} {\bibinfo
  {journal} {Phys. Rev. C}\ }\textbf {\bibinfo {volume} {80}},\ \bibinfo
  {pages} {025205} (\bibinfo {year} {2009})}\BibitemShut {NoStop}%
\bibitem [{\citenamefont {Nagahiro}\ \emph {et~al.}(2013)\citenamefont
  {Nagahiro}, \citenamefont {Jido}, \citenamefont {Fujioka}, \citenamefont
  {Itahashi},\ and\ \citenamefont {Hirenzaki}}]{NAGAHIRO2}%
  \BibitemOpen
  \bibfield  {author} {\bibinfo {author} {\bibfnamefont {H.}~\bibnamefont
  {Nagahiro}}, \bibinfo {author} {\bibfnamefont {D.}~\bibnamefont {Jido}},
  \bibinfo {author} {\bibfnamefont {H.}~\bibnamefont {Fujioka}}, \bibinfo
  {author} {\bibfnamefont {K.}~\bibnamefont {Itahashi}},\ and\ \bibinfo
  {author} {\bibfnamefont {S.}~\bibnamefont {Hirenzaki}},\ }\href
  {https://doi.org/10.1103/PhysRevC.87.045201} {\bibfield  {journal} {\bibinfo
  {journal} {Phys. Rev. C}\ }\textbf {\bibinfo {volume} {87}},\ \bibinfo
  {pages} {045201} (\bibinfo {year} {2013})}\BibitemShut {NoStop}%
\bibitem [{\citenamefont {Jido}\ \emph {et~al.}(2019)\citenamefont {Jido},
  \citenamefont {Masutani},\ and\ \citenamefont {Hirenzaki}}]{JIDO}%
  \BibitemOpen
  \bibfield  {author} {\bibinfo {author} {\bibfnamefont {D.}~\bibnamefont
  {Jido}}, \bibinfo {author} {\bibfnamefont {H.}~\bibnamefont {Masutani}},\
  and\ \bibinfo {author} {\bibfnamefont {S.}~\bibnamefont {Hirenzaki}},\ }\href
  {https://doi.org/10.1093/ptep/ptz031} {\bibfield  {journal} {\bibinfo
  {journal} {Progress of Theoretical and Experimental Physics}\ }\textbf
  {\bibinfo {volume} {2019}},\ \bibinfo {pages} {053D02} (\bibinfo {year}
  {2019})}\BibitemShut {NoStop}%
\bibitem [{\citenamefont {Tolos}\ and\ \citenamefont
  {Fabbietti}(2020)}]{TOLOS}%
  \BibitemOpen
  \bibfield  {author} {\bibinfo {author} {\bibfnamefont {L.}~\bibnamefont
  {Tolos}}\ and\ \bibinfo {author} {\bibfnamefont {L.}~\bibnamefont
  {Fabbietti}},\ }\href
  {https://doi.org/https://doi.org/10.1016/j.ppnp.2020.103770} {\bibfield
  {journal} {\bibinfo  {journal} {Progress in Particle and Nuclear Physics}\
  }\textbf {\bibinfo {volume} {112}},\ \bibinfo {pages} {103770} (\bibinfo
  {year} {2020})}\BibitemShut {NoStop}%
\bibitem [{\citenamefont {Glendenning}\ and\ \citenamefont
  {Schaffner-Bielich}(1998)}]{GLENDENNING2}%
  \BibitemOpen
  \bibfield  {author} {\bibinfo {author} {\bibfnamefont {N.~K.}\ \bibnamefont
  {Glendenning}}\ and\ \bibinfo {author} {\bibfnamefont {J.}~\bibnamefont
  {Schaffner-Bielich}},\ }\href {https://doi.org/10.1103/PhysRevLett.81.4564}
  {\bibfield  {journal} {\bibinfo  {journal} {Phys. Rev. Lett.}\ }\textbf
  {\bibinfo {volume} {81}},\ \bibinfo {pages} {4564} (\bibinfo {year}
  {1998})}\BibitemShut {NoStop}%
\bibitem [{\citenamefont {{Mishra}}\ \emph {et~al.}(2010)\citenamefont
  {{Mishra}}, \citenamefont {{Kumar}}, \citenamefont {{Sanyal}}, \citenamefont
  {{Dexheimer}},\ and\ \citenamefont {{Schramm}}}]{MISHRA_SANYAL}%
  \BibitemOpen
  \bibfield  {author} {\bibinfo {author} {\bibfnamefont {A.}~\bibnamefont
  {{Mishra}}}, \bibinfo {author} {\bibfnamefont {A.}~\bibnamefont {{Kumar}}},
  \bibinfo {author} {\bibfnamefont {S.}~\bibnamefont {{Sanyal}}}, \bibinfo
  {author} {\bibfnamefont {V.}~\bibnamefont {{Dexheimer}}},\ and\ \bibinfo
  {author} {\bibfnamefont {S.}~\bibnamefont {{Schramm}}},\ }\href
  {https://doi.org/10.1140/epja/i2010-10986-x} {\bibfield  {journal} {\bibinfo
  {journal} {European Physical Journal A}\ }\textbf {\bibinfo {volume} {45}},\
  \bibinfo {pages} {169} (\bibinfo {year} {2010})}\BibitemShut {NoStop}%
\bibitem [{\citenamefont {Maruyama}\ \emph {et~al.}(2005)\citenamefont
  {Maruyama}, \citenamefont {Muto}, \citenamefont {Tatsumi}, \citenamefont
  {Tsushima},\ and\ \citenamefont {Thomas}}]{MARUYAMA}%
  \BibitemOpen
  \bibfield  {author} {\bibinfo {author} {\bibfnamefont {T.}~\bibnamefont
  {Maruyama}}, \bibinfo {author} {\bibfnamefont {T.}~\bibnamefont {Muto}},
  \bibinfo {author} {\bibfnamefont {T.}~\bibnamefont {Tatsumi}}, \bibinfo
  {author} {\bibfnamefont {K.}~\bibnamefont {Tsushima}},\ and\ \bibinfo
  {author} {\bibfnamefont {A.~W.}\ \bibnamefont {Thomas}},\ }\href
  {https://doi.org/https://doi.org/10.1016/j.nuclphysa.2005.06.008} {\bibfield
  {journal} {\bibinfo  {journal} {Nuclear Physics A}\ }\textbf {\bibinfo
  {volume} {760}},\ \bibinfo {pages} {319} (\bibinfo {year}
  {2005})}\BibitemShut {NoStop}%
\bibitem [{\citenamefont {Zhong}\ \emph {et~al.}(2006)\citenamefont {Zhong},
  \citenamefont {Peng}, \citenamefont {Li},\ and\ \citenamefont
  {Ning}}]{ZHONG}%
  \BibitemOpen
  \bibfield  {author} {\bibinfo {author} {\bibfnamefont {X.~H.}\ \bibnamefont
  {Zhong}}, \bibinfo {author} {\bibfnamefont {G.~X.}\ \bibnamefont {Peng}},
  \bibinfo {author} {\bibfnamefont {L.}~\bibnamefont {Li}},\ and\ \bibinfo
  {author} {\bibfnamefont {P.~Z.}\ \bibnamefont {Ning}},\ }\href
  {https://doi.org/10.1103/PhysRevC.73.015205} {\bibfield  {journal} {\bibinfo
  {journal} {Phys. Rev. C}\ }\textbf {\bibinfo {volume} {73}},\ \bibinfo
  {pages} {015205} (\bibinfo {year} {2006})}\BibitemShut {NoStop}%
\bibitem [{\citenamefont {Kumar}\ and\ \citenamefont {Kumar}(2020)}]{KUMAR}%
  \BibitemOpen
  \bibfield  {author} {\bibinfo {author} {\bibfnamefont {R.}~\bibnamefont
  {Kumar}}\ and\ \bibinfo {author} {\bibfnamefont {A.}~\bibnamefont {Kumar}},\
  }\href {https://doi.org/10.1103/PhysRevC.102.065207} {\bibfield  {journal}
  {\bibinfo  {journal} {Phys. Rev. C}\ }\textbf {\bibinfo {volume} {102}},\
  \bibinfo {pages} {065207} (\bibinfo {year} {2020})}\BibitemShut {NoStop}%
\bibitem [{\citenamefont {Ratti}\ \emph {et~al.}(2006)\citenamefont {Ratti},
  \citenamefont {Thaler},\ and\ \citenamefont {Weise}}]{ratti_weise}%
  \BibitemOpen
  \bibfield  {author} {\bibinfo {author} {\bibfnamefont {C.}~\bibnamefont
  {Ratti}}, \bibinfo {author} {\bibfnamefont {M.~A.}\ \bibnamefont {Thaler}},\
  and\ \bibinfo {author} {\bibfnamefont {W.}~\bibnamefont {Weise}},\ }\href
  {https://doi.org/10.1103/PhysRevD.73.014019} {\bibfield  {journal} {\bibinfo
  {journal} {Phys. Rev. D}\ }\textbf {\bibinfo {volume} {73}},\ \bibinfo
  {pages} {014019} (\bibinfo {year} {2006})}\BibitemShut {NoStop}%
\bibitem [{\citenamefont {{Ratti}}\ \emph {et~al.}(2007)\citenamefont
  {{Ratti}}, \citenamefont {{R{\"o}{\ss}ner}}, \citenamefont {{Thaler}},\ and\
  \citenamefont {{Weise}}}]{ratti_weise2}%
  \BibitemOpen
  \bibfield  {author} {\bibinfo {author} {\bibfnamefont {C.}~\bibnamefont
  {{Ratti}}}, \bibinfo {author} {\bibfnamefont {S.}~\bibnamefont
  {{R{\"o}{\ss}ner}}}, \bibinfo {author} {\bibfnamefont {M.~A.}\ \bibnamefont
  {{Thaler}}},\ and\ \bibinfo {author} {\bibfnamefont {W.}~\bibnamefont
  {{Weise}}},\ }\href {https://doi.org/10.1140/epjc/s10052-006-0065-x}
  {\bibfield  {journal} {\bibinfo  {journal} {European Physical Journal C}\
  }\textbf {\bibinfo {volume} {49}},\ \bibinfo {pages} {213} (\bibinfo {year}
  {2007})}\BibitemShut {NoStop}%
\bibitem [{\citenamefont {Fukushima}(2008)}]{fukushima}%
  \BibitemOpen
  \bibfield  {author} {\bibinfo {author} {\bibfnamefont {K.}~\bibnamefont
  {Fukushima}},\ }\href {https://doi.org/10.1103/PhysRevD.77.114028} {\bibfield
   {journal} {\bibinfo  {journal} {Phys. Rev. D}\ }\textbf {\bibinfo {volume}
  {77}},\ \bibinfo {pages} {114028} (\bibinfo {year} {2008})}\BibitemShut
  {NoStop}%
\bibitem [{\citenamefont {R\"o\ss{}ner}\ \emph {et~al.}(2007)\citenamefont
  {R\"o\ss{}ner}, \citenamefont {Ratti},\ and\ \citenamefont
  {Weise}}]{rossner}%
  \BibitemOpen
  \bibfield  {author} {\bibinfo {author} {\bibfnamefont {S.}~\bibnamefont
  {R\"o\ss{}ner}}, \bibinfo {author} {\bibfnamefont {C.}~\bibnamefont
  {Ratti}},\ and\ \bibinfo {author} {\bibfnamefont {W.}~\bibnamefont {Weise}},\
  }\href {https://doi.org/10.1103/PhysRevD.75.034007} {\bibfield  {journal}
  {\bibinfo  {journal} {Phys. Rev. D}\ }\textbf {\bibinfo {volume} {75}},\
  \bibinfo {pages} {034007} (\bibinfo {year} {2007})}\BibitemShut {NoStop}%
\bibitem [{\citenamefont {Gaitanos}\ \emph {et~al.}(2004)\citenamefont
  {Gaitanos}, \citenamefont {{Di Toro}}, \citenamefont {Typel}, \citenamefont
  {Baran}, \citenamefont {Fuchs}, \citenamefont {Greco},\ and\ \citenamefont
  {Wolter}}]{gaitanos}%
  \BibitemOpen
  \bibfield  {author} {\bibinfo {author} {\bibfnamefont {T.}~\bibnamefont
  {Gaitanos}}, \bibinfo {author} {\bibfnamefont {M.}~\bibnamefont {{Di Toro}}},
  \bibinfo {author} {\bibfnamefont {S.}~\bibnamefont {Typel}}, \bibinfo
  {author} {\bibfnamefont {V.}~\bibnamefont {Baran}}, \bibinfo {author}
  {\bibfnamefont {C.}~\bibnamefont {Fuchs}}, \bibinfo {author} {\bibfnamefont
  {V.}~\bibnamefont {Greco}},\ and\ \bibinfo {author} {\bibfnamefont
  {H.}~\bibnamefont {Wolter}},\ }\href
  {https://doi.org/https://doi.org/10.1016/j.nuclphysa.2003.12.001} {\bibfield
  {journal} {\bibinfo  {journal} {Nuclear Physics A}\ }\textbf {\bibinfo
  {volume} {732}},\ \bibinfo {pages} {24} (\bibinfo {year} {2004})}\BibitemShut
  {NoStop}%
\bibitem [{\citenamefont {Avancini}\ \emph {et~al.}(2009)\citenamefont
  {Avancini}, \citenamefont {Brito}, \citenamefont {Marinelli}, \citenamefont
  {Menezes}, \citenamefont {de~Moraes}, \citenamefont {Provid\^encia},\ and\
  \citenamefont {Santos}}]{debora2009}%
  \BibitemOpen
  \bibfield  {author} {\bibinfo {author} {\bibfnamefont {S.~S.}\ \bibnamefont
  {Avancini}}, \bibinfo {author} {\bibfnamefont {L.}~\bibnamefont {Brito}},
  \bibinfo {author} {\bibfnamefont {J.~R.}\ \bibnamefont {Marinelli}}, \bibinfo
  {author} {\bibfnamefont {D.~P.}\ \bibnamefont {Menezes}}, \bibinfo {author}
  {\bibfnamefont {M.~M.~W.}\ \bibnamefont {de~Moraes}}, \bibinfo {author}
  {\bibfnamefont {C.}~\bibnamefont {Provid\^encia}},\ and\ \bibinfo {author}
  {\bibfnamefont {A.~M.}\ \bibnamefont {Santos}},\ }\href
  {https://doi.org/10.1103/PhysRevC.79.035804} {\bibfield  {journal} {\bibinfo
  {journal} {Phys. Rev. C}\ }\textbf {\bibinfo {volume} {79}},\ \bibinfo
  {pages} {035804} (\bibinfo {year} {2009})}\BibitemShut {NoStop}%
\bibitem [{\citenamefont {Fortin}\ \emph {et~al.}(2016)\citenamefont {Fortin},
  \citenamefont {Provid\^encia}, \citenamefont {Raduta}, \citenamefont
  {Gulminelli}, \citenamefont {Zdunik}, \citenamefont {Haensel},\ and\
  \citenamefont {Bejger}}]{fortin}%
  \BibitemOpen
  \bibfield  {author} {\bibinfo {author} {\bibfnamefont {M.}~\bibnamefont
  {Fortin}}, \bibinfo {author} {\bibfnamefont {C.}~\bibnamefont
  {Provid\^encia}}, \bibinfo {author} {\bibfnamefont {A.~R.}\ \bibnamefont
  {Raduta}}, \bibinfo {author} {\bibfnamefont {F.}~\bibnamefont {Gulminelli}},
  \bibinfo {author} {\bibfnamefont {J.~L.}\ \bibnamefont {Zdunik}}, \bibinfo
  {author} {\bibfnamefont {P.}~\bibnamefont {Haensel}},\ and\ \bibinfo {author}
  {\bibfnamefont {M.}~\bibnamefont {Bejger}},\ }\href
  {https://doi.org/10.1103/PhysRevC.94.035804} {\bibfield  {journal} {\bibinfo
  {journal} {Phys. Rev. C}\ }\textbf {\bibinfo {volume} {94}},\ \bibinfo
  {pages} {035804} (\bibinfo {year} {2016})}\BibitemShut {NoStop}%
\bibitem [{\citenamefont {Klevansky}(1992)}]{KLEVANSKY}%
  \BibitemOpen
  \bibfield  {author} {\bibinfo {author} {\bibfnamefont {S.~P.}\ \bibnamefont
  {Klevansky}},\ }\href {https://doi.org/10.1103/RevModPhys.64.649} {\bibfield
  {journal} {\bibinfo  {journal} {Rev. Mod. Phys.}\ }\textbf {\bibinfo {volume}
  {64}},\ \bibinfo {pages} {649} (\bibinfo {year} {1992})}\BibitemShut
  {NoStop}%
\bibitem [{\citenamefont {Rehberg}\ \emph {et~al.}(1996)\citenamefont
  {Rehberg}, \citenamefont {Klevansky},\ and\ \citenamefont
  {H\"ufner}}]{REHBERG}%
  \BibitemOpen
  \bibfield  {author} {\bibinfo {author} {\bibfnamefont {P.}~\bibnamefont
  {Rehberg}}, \bibinfo {author} {\bibfnamefont {S.~P.}\ \bibnamefont
  {Klevansky}},\ and\ \bibinfo {author} {\bibfnamefont {J.}~\bibnamefont
  {H\"ufner}},\ }\href {https://doi.org/10.1103/PhysRevC.53.410} {\bibfield
  {journal} {\bibinfo  {journal} {Phys. Rev. C}\ }\textbf {\bibinfo {volume}
  {53}},\ \bibinfo {pages} {410} (\bibinfo {year} {1996})}\BibitemShut
  {NoStop}%
\bibitem [{\citenamefont {Costa}\ \emph {et~al.}(2004)\citenamefont {Costa},
  \citenamefont {Ruivo}, \citenamefont {de~Sousa},\ and\ \citenamefont
  {Kalinovsky}}]{COSTA}%
  \BibitemOpen
  \bibfield  {author} {\bibinfo {author} {\bibfnamefont {P.}~\bibnamefont
  {Costa}}, \bibinfo {author} {\bibfnamefont {M.~C.}\ \bibnamefont {Ruivo}},
  \bibinfo {author} {\bibfnamefont {C.~A.}\ \bibnamefont {de~Sousa}},\ and\
  \bibinfo {author} {\bibfnamefont {Y.~L.}\ \bibnamefont {Kalinovsky}},\ }\href
  {https://doi.org/10.1103/PhysRevC.70.025204} {\bibfield  {journal} {\bibinfo
  {journal} {Phys. Rev. C}\ }\textbf {\bibinfo {volume} {70}},\ \bibinfo
  {pages} {025204} (\bibinfo {year} {2004})}\BibitemShut {NoStop}%
\bibitem [{\citenamefont {Brown}\ \emph {et~al.}(1991)\citenamefont {Brown},
  \citenamefont {Koch},\ and\ \citenamefont {Rho}}]{BROWN_RHO2}%
  \BibitemOpen
  \bibfield  {author} {\bibinfo {author} {\bibfnamefont {G.}~\bibnamefont
  {Brown}}, \bibinfo {author} {\bibfnamefont {V.}~\bibnamefont {Koch}},\ and\
  \bibinfo {author} {\bibfnamefont {M.}~\bibnamefont {Rho}},\ }\href
  {https://doi.org/https://doi.org/10.1016/0375-9474(91)90483-M} {\bibfield
  {journal} {\bibinfo  {journal} {Nuclear Physics A}\ }\textbf {\bibinfo
  {volume} {535}},\ \bibinfo {pages} {701} (\bibinfo {year}
  {1991})}\BibitemShut {NoStop}%
\bibitem [{\citenamefont {Rapp}\ \emph {et~al.}(1999)\citenamefont {Rapp},
  \citenamefont {Machleidt}, \citenamefont {Durso},\ and\ \citenamefont
  {Brown}}]{RAPP}%
  \BibitemOpen
  \bibfield  {author} {\bibinfo {author} {\bibfnamefont {R.}~\bibnamefont
  {Rapp}}, \bibinfo {author} {\bibfnamefont {R.}~\bibnamefont {Machleidt}},
  \bibinfo {author} {\bibfnamefont {J.~W.}\ \bibnamefont {Durso}},\ and\
  \bibinfo {author} {\bibfnamefont {G.~E.}\ \bibnamefont {Brown}},\ }\href
  {https://doi.org/10.1103/PhysRevLett.82.1827} {\bibfield  {journal} {\bibinfo
   {journal} {Phys. Rev. Lett.}\ }\textbf {\bibinfo {volume} {82}},\ \bibinfo
  {pages} {1827} (\bibinfo {year} {1999})}\BibitemShut {NoStop}%
\bibitem [{\citenamefont {Brown}\ and\ \citenamefont {Rho}(1991)}]{BROWN_RHO}%
  \BibitemOpen
  \bibfield  {author} {\bibinfo {author} {\bibfnamefont {G.~E.}\ \bibnamefont
  {Brown}}\ and\ \bibinfo {author} {\bibfnamefont {M.}~\bibnamefont {Rho}},\
  }\href {https://doi.org/10.1103/PhysRevLett.66.2720} {\bibfield  {journal}
  {\bibinfo  {journal} {Phys. Rev. Lett.}\ }\textbf {\bibinfo {volume} {66}},\
  \bibinfo {pages} {2720} (\bibinfo {year} {1991})}\BibitemShut {NoStop}%
\bibitem [{\citenamefont {Saito}\ \emph {et~al.}(1989)\citenamefont {Saito},
  \citenamefont {Maruyama},\ and\ \citenamefont {Soutome}}]{SAITO}%
  \BibitemOpen
  \bibfield  {author} {\bibinfo {author} {\bibfnamefont {K.}~\bibnamefont
  {Saito}}, \bibinfo {author} {\bibfnamefont {T.}~\bibnamefont {Maruyama}},\
  and\ \bibinfo {author} {\bibfnamefont {K.}~\bibnamefont {Soutome}},\ }\href
  {https://doi.org/10.1103/PhysRevC.40.407} {\bibfield  {journal} {\bibinfo
  {journal} {Phys. Rev. C}\ }\textbf {\bibinfo {volume} {40}},\ \bibinfo
  {pages} {407} (\bibinfo {year} {1989})}\BibitemShut {NoStop}%
\bibitem [{\citenamefont {Kaplan}\ and\ \citenamefont {Nelson}(1986)}]{KAPLAN}%
  \BibitemOpen
  \bibfield  {author} {\bibinfo {author} {\bibfnamefont {D.}~\bibnamefont
  {Kaplan}}\ and\ \bibinfo {author} {\bibfnamefont {A.}~\bibnamefont
  {Nelson}},\ }\href
  {https://doi.org/https://doi.org/10.1016/0370-2693(86)90331-X} {\bibfield
  {journal} {\bibinfo  {journal} {Physics Letters B}\ }\textbf {\bibinfo
  {volume} {175}},\ \bibinfo {pages} {57} (\bibinfo {year} {1986})}\BibitemShut
  {NoStop}%
\bibitem [{\citenamefont {Brown}\ \emph {et~al.}(1994)\citenamefont {Brown},
  \citenamefont {Lee}, \citenamefont {Rho},\ and\ \citenamefont
  {Thorsson}}]{BROWN_LEE}%
  \BibitemOpen
  \bibfield  {author} {\bibinfo {author} {\bibfnamefont {G.}~\bibnamefont
  {Brown}}, \bibinfo {author} {\bibfnamefont {C.-H.}\ \bibnamefont {Lee}},
  \bibinfo {author} {\bibfnamefont {M.}~\bibnamefont {Rho}},\ and\ \bibinfo
  {author} {\bibfnamefont {V.}~\bibnamefont {Thorsson}},\ }\href
  {https://doi.org/https://doi.org/10.1016/0375-9474(94)90335-2} {\bibfield
  {journal} {\bibinfo  {journal} {Nuclear Physics A}\ }\textbf {\bibinfo
  {volume} {567}},\ \bibinfo {pages} {937} (\bibinfo {year}
  {1994})}\BibitemShut {NoStop}%
\bibitem [{\citenamefont {Weinberg}(1967)}]{WEINBERG}%
  \BibitemOpen
  \bibfield  {author} {\bibinfo {author} {\bibfnamefont {S.}~\bibnamefont
  {Weinberg}},\ }\href {https://doi.org/10.1103/PhysRevLett.18.188} {\bibfield
  {journal} {\bibinfo  {journal} {Phys. Rev. Lett.}\ }\textbf {\bibinfo
  {volume} {18}},\ \bibinfo {pages} {188} (\bibinfo {year} {1967})}\BibitemShut
  {NoStop}%
\bibitem [{\citenamefont {Papazoglou}\ \emph {et~al.}(1998)\citenamefont
  {Papazoglou}, \citenamefont {Schramm}, \citenamefont {Schaffner-Bielich},
  \citenamefont {St\"ocker},\ and\ \citenamefont {Greiner}}]{PAPAZOGLOU}%
  \BibitemOpen
  \bibfield  {author} {\bibinfo {author} {\bibfnamefont {P.}~\bibnamefont
  {Papazoglou}}, \bibinfo {author} {\bibfnamefont {S.}~\bibnamefont {Schramm}},
  \bibinfo {author} {\bibfnamefont {J.}~\bibnamefont {Schaffner-Bielich}},
  \bibinfo {author} {\bibfnamefont {H.}~\bibnamefont {St\"ocker}},\ and\
  \bibinfo {author} {\bibfnamefont {W.}~\bibnamefont {Greiner}},\ }\href
  {https://doi.org/10.1103/PhysRevC.57.2576} {\bibfield  {journal} {\bibinfo
  {journal} {Phys. Rev. C}\ }\textbf {\bibinfo {volume} {57}},\ \bibinfo
  {pages} {2576} (\bibinfo {year} {1998})}\BibitemShut {NoStop}%
\bibitem [{\citenamefont {Costa}\ and\ \citenamefont {Pereira}(2019)}]{COSTA3}%
  \BibitemOpen
  \bibfield  {author} {\bibinfo {author} {\bibfnamefont {P.}~\bibnamefont
  {Costa}}\ and\ \bibinfo {author} {\bibfnamefont {R.~C.}\ \bibnamefont
  {Pereira}},\ }\href {https://doi.org/10.3390/sym11040507} {\bibfield
  {journal} {\bibinfo  {journal} {Symmetry}\ }\textbf {\bibinfo {volume}
  {11}},\ \bibinfo {pages} {507} (\bibinfo {year} {2019})},\ \Eprint
  {https://arxiv.org/abs/1904.05805} {arXiv:1904.05805 [hep-ph]} \BibitemShut
  {NoStop}%
\bibitem [{\citenamefont {Bernard}\ and\ \citenamefont
  {Meissner}(1988)}]{BERNARD}%
  \BibitemOpen
  \bibfield  {author} {\bibinfo {author} {\bibfnamefont {V.}~\bibnamefont
  {Bernard}}\ and\ \bibinfo {author} {\bibfnamefont {U.~G.}\ \bibnamefont
  {Meissner}},\ }\href {https://doi.org/10.1103/PhysRevD.38.1551} {\bibfield
  {journal} {\bibinfo  {journal} {Phys. Rev. D}\ }\textbf {\bibinfo {volume}
  {38}},\ \bibinfo {pages} {1551} (\bibinfo {year} {1988})}\BibitemShut
  {NoStop}%
\bibitem [{\citenamefont {de~Sousa}\ and\ \citenamefont {Ruivo}(1997)}]{SOUSA}%
  \BibitemOpen
  \bibfield  {author} {\bibinfo {author} {\bibfnamefont {C.~A.}\ \bibnamefont
  {de~Sousa}}\ and\ \bibinfo {author} {\bibfnamefont {M.~C.}\ \bibnamefont
  {Ruivo}},\ }\href {https://doi.org/10.1016/S0375-9474(97)00389-8} {\bibfield
  {journal} {\bibinfo  {journal} {Nucl. Phys. A}\ }\textbf {\bibinfo {volume}
  {625}},\ \bibinfo {pages} {713} (\bibinfo {year} {1997})}\BibitemShut
  {NoStop}%
\end{thebibliography}%

\end{document}